\documentclass[a4paper,11pt]{article}
\pdfoutput=1 % if your are submitting a pdflatex (i.e. if you have
\usepackage{jheppub} % for details on the use of the package, please\eta
\usepackage{amssymb} 
\usepackage{amsmath}
\usepackage{mathtools}
\usepackage{amsfonts}    
\usepackage{dsfont}
\usepackage{pdfpages}
\usepackage{verbatim}
\usepackage{tensor}
\usepackage{mathrsfs}
\usepackage[mathscr]{euscript}
\usepackage{tikz}
\usetikzlibrary{calc}
\usepackage{braket}
\usepackage{hyperref}

\usetikzlibrary{shapes.misc}
\tikzset{cross/.style={cross out, draw=black, ultra thick, minimum size=2*(#1-\pgflinewidth), inner sep=0pt, outer sep=0pt},
cross/.default={5pt}}

\let\oldbibliography\thebibliography
\renewcommand{\thebibliography}[1]{\oldbibliography{#1}
\setlength{\itemsep}{4.14pt}} %Reducing spacing in the bibliography.

\allowdisplaybreaks
\usepackage{mathrsfs}
\usepackage[mathscr]{euscript}

\numberwithin{equation}{section}

\def\be{\begin{equation}}
\def\ee{\end{equation}}

\title{\boldmath From Symmetric Product CFTs to ${\rm AdS}_3$}

\author[a]{\!\! Matthias R.~Gaberdiel}
\author[b]{\!\!, Rajesh Gopakumar}
\author[a]{\!\!, Bob Knighton}
\author[b]{\!\! and Pronobesh Maity}

\affiliation[a]{Institut f\"ur Theoretische Physik, ETH Z\"urich \\
\hspace*{0.3cm} Wolfgang-Pauli-Stra{\ss}e 27, 8093 Z\"urich, Switzerland}
\affiliation[b]{International Centre for Theoretical Sciences-TIFR, \\
\hspace*{0.3cm} Shivakote, Hesaraghatta Hobli, Bengaluru North, India 560 089}

\emailAdd{gaberdiel@itp.phys.ethz.ch}
\emailAdd{rajesh.gopakumar@icts.res.in}
\emailAdd{robejr@ethz.ch}
\emailAdd{pronobesh.maity@icts.res.in}

\abstract{Correlators in symmetric orbifold CFTs are given by a finite sum of admissible branched covers of the 2d spacetime. We consider a Gross-Mende like limit where all operators have large twist, and show that the corresponding branched covers can be described via a Penner-like matrix model. The limiting branched covers are given in terms of the spectral curve for this matrix model, which remarkably turns out to be directly related to the Strebel quadratic differential on the covering space. Interpreting the covering space as the world-sheet of the dual string theory, the spacetime CFT correlator  thus has the form of an integral over the entire world-sheet moduli space weighted with a Nambu-Goto-like action. Quite strikingly, at leading order this action can also be written  as the absolute value of the Schwarzian of the covering map. 

Given the equivalence of the symmetric product CFT to tensionless string theory on ${\rm AdS}_3$, this provides an explicit realisation of the underlying mechanism of gauge-string duality  originally proposed in \cite{Gopakumar:2005fx} and further refined in \cite{Razamat:2008zr}.}

\begin{document}
\maketitle
\flushbottom

\section{Introduction}

How exactly do quantum field theories reassemble themselves into string theories (or M-theory generalisations) in the large $N$ limit? This question has been with us ever since 't~Hooft showed that Feynman diagrams of large $N$ gauge theories admit a genus expansion \cite{tHooft:1973alw} which hinted at a dual string theory description. Maldacena's discovery \cite{Maldacena:1997re}, through the physics of D-branes, of the string dual for a large class of supersymmetric gauge theories gave this question fresh impetus.  It provided us with a set of quantum field theories where one had a precise dictionary \cite{Gubser:1998bc, Witten:1998qj} between observables, to those of a dual string theory, with the latter often admitting a weakly coupled gravity description. 

More specifically, the duality gives rise to a map between 
(single trace) gauge invariant operators, and vertex operators for perturbative string states ($w$ and $h$ label the state, while $z$ is a worldsheet coordinate)
\be
{\cal O}^{(w)}_{h}(x) \quad \longleftrightarrow  \quad \mathcal{V}^{w}_{h}(x; z) \ . 
\ee
The relation then between $n$-point (Euclidean) correlators is, 
\begin{multline} \label{corresp}
\big\langle {\cal O}^{(w_1)}_{h_1}(x_1){\cal O}^{(w_2)}_{h_2}(x_2) \ldots {\cal O}^{(w_n)}_{h_n}(x_n) \big\rangle_{\text{S}^d} \Big|_{g} \\ 
= \int_{{\cal M}_{g,n}} \!\!\! \! \!\! d\mu \, \big\langle \mathcal{V}^{w_1}_{h_1}(x_1; z_1)\mathcal{V}^{w_2}_{h_2}(x_2; z_2) \ldots \mathcal{V}_{h_n}^{w_n}(x_n; z_n) \big\rangle_{\Sigma_{g,n}}    \ ,
\end{multline}
where $g$ is the genus of the world-sheet, while on the CFT side (i.e.\ the LHS) it captures a certain contribution in the $\frac{1}{N}$ expansion. Both sides of this equality are autonomously defined, and it should be possible to decipher the mechanism behind this remarkable correspondence. However, up to now this question has remained unanswered in any precise sense --- the miraculous nature of the equality between the LHS and RHS of (\ref{corresp}) has remained so  even after two decades. 

Some time ago, a proposal was made  for what this underlying mechanism might be \cite{Gopakumar:2003ns, Gopakumar:2004qb,Gopakumar:2005fx}, in an expansion around the free field point on the LHS. This weak coupling limit of the quantum field theory 
translates into a tensionless (or high curvature) limit of the dual string theory, and this is also the regime of the Feynman diagram expansion which underlies the original 't~Hooft analysis. The basic idea behind the proposal of 
\cite{Gopakumar:2003ns, Gopakumar:2004qb,Gopakumar:2005fx} was to reorganise the sum over distinct worldline trajectories that Feynman diagrams represent, into a sum over distinct world-sheets which are glued up versions of the original double line graphs.  In other words, it is a prescriptive procedure for how to go from the LHS to the RHS of (\ref{corresp}). Note that this is a reverse engineering problem in that one aims to reconstruct an integrand on moduli space, and there is {\it a priori} no unique way of doing so. That does not, however, obviate the possibility of a canonical or natural way of going between the two sides of (\ref{corresp}), and indeed the proposal of \cite{Gopakumar:2003ns, Gopakumar:2004qb,Gopakumar:2005fx} gives rise to a particular integrand on the RHS. 

The first step in this reorganisation was to group together genus $g$ Feynman diagrams in the free theory that contribute to a gauge invariant $n$-point correlator 
--- the LHS of (\ref{corresp}) --- into `skeleton graphs'  \cite{Gopakumar:2004qb}. This essentially involved gluing together homotopically equivalent edges of the Feynman graph. These skeleton graphs capture, in a sense, the inequivalent topologies of the worldlines. It was then seen that these skeleton graphs are precisely in correspondence with a simplicial decomposition of the (decorated) moduli space ${\cal M}_{g,n}\times {\mathbb R}_+^n$. This can be thought of as a refinement of 't~Hooft's association of a genus to Feynman diagrams. The next step  \cite{Gopakumar:2005fx} was to make a one-to-one correspondence between individual diagrams and individual world-sheets, i.e.\ the points on the moduli space on the RHS of (\ref{corresp}). 
This exploited the mathematics of Strebel differentials which underlies the above cell decomposition of moduli space. 

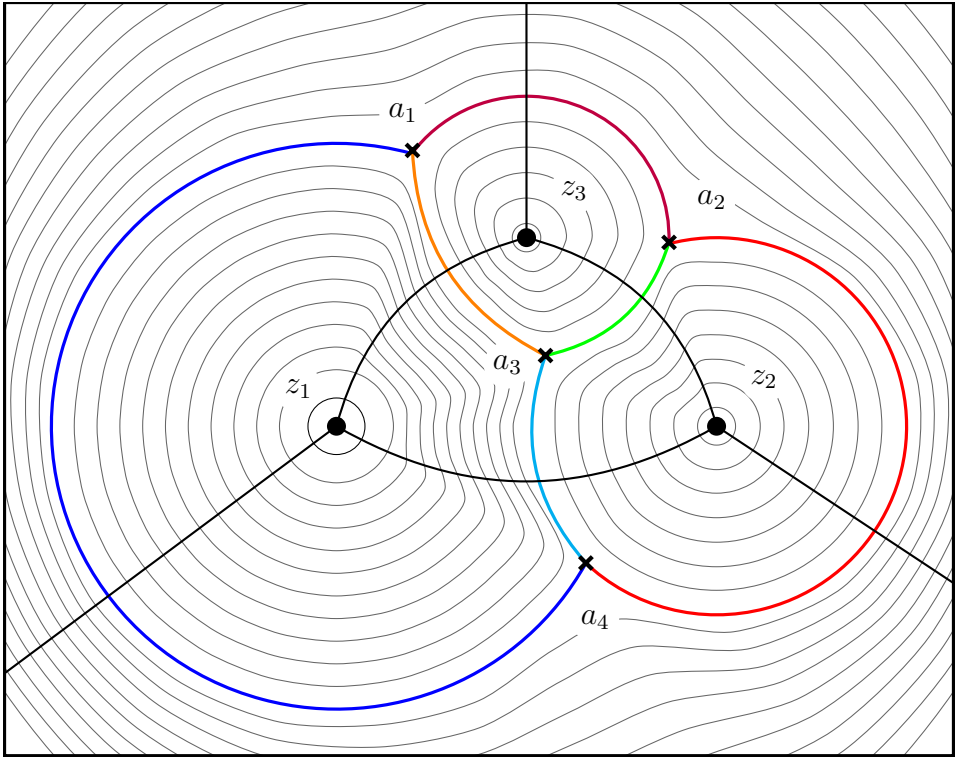
\begin{figure}
\centering
\begin{tikzpicture}[scale = 1.25]

%defining the center of the picture

\coordinate (W) at (1.5,0.5);

% clipping the whole thing

\clip ($(W) + (5,4)$) -- ($(W) + (5,-4)$) -- ($(W) + (-5,-4)$) -- ($(W) + (-5,4)$) -- ($(W) + (5,4)$);
\draw[ultra thick] ($(W) + (5,4)$) -- ($(W) + (5,-4)$) -- ($(W) + (-5,-4)$) -- ($(W) + (-5,4)$) -- ($(W) + (5,4)$);

%defining the coordinates of the important points -- A,B,C,D are the intersections of the circles, whereas X,Y,Z are the three insertion points (with the fourth at infinity)

\coordinate (A) at (0.8,2.925);
\coordinate (B) at (3.5,1.95);
\coordinate (C) at (2.2,0.75);
\coordinate (D) at (2.625,-1.45);
\coordinate (X) at (0,0);
\coordinate (Y) at (4,0);
\coordinate (Z) at (2,2);

%drawing the circles that define borders of the equal-time trajectories

\draw[very thick, blue] ([shift = (75:3cm)]X) arc (75:331:3cm);
\draw[very thick, red] ([shift = (105:2cm)]Y) arc (105:-134:2cm);
\draw[very thick, purple] ([shift = (-2:1.5cm)]Z) arc (-2:144:1.5cm);
\draw[very thick, orange] (A) to[bend right] (C);
\draw[very thick, cyan] (C) to[bend right] (D);
\draw[very thick, green] (B) to[bend left] (C);

%insertion points and intersections get some volume

\draw[red] (A) node[cross = 4, color = black, rotate = 0]{};
\draw[red] (B) node[cross = 4, color = black, rotate = 0]{};
\draw[red] (C) node[cross = 4, color = black, rotate = 0]{};
\draw[red] (D) node[cross = 4, color = black, rotate = 0]{};
\fill (X) circle (0.1);
\fill (Y) circle (0.1);
\fill (Z) circle (0.1);

%graph connecting insertion points

\draw[thick] (X) to[out = -30, in = -150] (Y) to[out = 135-30, in = -45+30] (Z) to[out = -135 - 30, in = 45 + 30] (X);
\draw[thick] (X) -- (-4,-3);
\draw[thick] (Y) -- (7,-2);
\draw[thick] (Z) -- (2,4.65);

%the hard part -- the equal-time contours

%the contours around the point (X)

\draw (X) circle (0.3);
\draw[opacity = 0.6] plot [smooth cycle, tension = 0.5] coordinates{($(X) + (0:1.86cm) $) ($(X) + (15:2.04cm) $) ($(X) + (30:1.86cm) $) ($(X) + (45:1.77cm) $) ($(X) + (60:1.95cm) $) ($(X) + (75:2.67cm) $) ($(X) + (90:2.76cm) $) ($(X) + (105:2.76cm) $) ($(X) + (120:2.76cm) $) ($(X) + (135:2.76cm) $) ($(X) + (150:2.76cm) $) ($(X) + (165:2.76cm) $) ($(X) + (180:2.76cm) $) ($(X) + (195:2.76cm) $) ($(X) + (210:2.76cm) $) ($(X) + (225:2.76cm) $) ($(X) + (240:2.76cm) $) ($(X) + (255:2.76cm) $) ($(X) + (270:2.76cm) $) ($(X) + (285:2.76cm) $) ($(X) + (300:2.76cm) $) ($(X) + (315:2.76cm) $) ($(X) + (330:2.76cm) $) ($(X) + (345:2.04cm) $) };
\draw[opacity = 0.6] plot [smooth cycle, tension = 0.5] coordinates{($(X) + (0:1.72cm) $) ($(X) + (15:1.88cm) $) ($(X) + (30:1.72cm) $) ($(X) + (45:1.64cm) $) ($(X) + (60:1.8cm) $) ($(X) + (75:2.44cm) $) ($(X) + (90:2.52cm) $) ($(X) + (105:2.52cm) $) ($(X) + (120:2.52cm) $) ($(X) + (135:2.52cm) $) ($(X) + (150:2.52cm) $) ($(X) + (165:2.52cm) $) ($(X) + (180:2.52cm) $) ($(X) + (195:2.52cm) $) ($(X) + (210:2.52cm) $) ($(X) + (225:2.52cm) $) ($(X) + (240:2.52cm) $) ($(X) + (255:2.52cm) $) ($(X) + (270:2.52cm) $) ($(X) + (285:2.52cm) $) ($(X) + (300:2.52cm) $) ($(X) + (315:2.52cm) $) ($(X) + (330:2.52cm) $) ($(X) + (345:1.88cm) $) };
\draw[opacity = 0.6] plot [smooth cycle, tension = 0.5] coordinates{($(X) + (0:1.58cm) $) ($(X) + (15:1.72cm) $) ($(X) + (30:1.58cm) $) ($(X) + (45:1.51cm) $) ($(X) + (60:1.65cm) $) ($(X) + (75:2.21cm) $) ($(X) + (90:2.28cm) $) ($(X) + (105:2.28cm) $) ($(X) + (120:2.28cm) $) ($(X) + (135:2.28cm) $) ($(X) + (150:2.28cm) $) ($(X) + (165:2.28cm) $) ($(X) + (180:2.28cm) $) ($(X) + (195:2.28cm) $) ($(X) + (210:2.28cm) $) ($(X) + (225:2.28cm) $) ($(X) + (240:2.28cm) $) ($(X) + (255:2.28cm) $) ($(X) + (270:2.28cm) $) ($(X) + (285:2.28cm) $) ($(X) + (300:2.28cm) $) ($(X) + (315:2.28cm) $) ($(X) + (330:2.28cm) $) ($(X) + (345:1.72cm) $) };
\draw[opacity = 0.6] plot [smooth cycle, tension = 0.5] coordinates{($(X) + (0:1.44cm) $) ($(X) + (15:1.56cm) $) ($(X) + (30:1.44cm) $) ($(X) + (45:1.38cm) $) ($(X) + (60:1.5cm) $) ($(X) + (75:1.98cm) $) ($(X) + (90:2.04cm) $) ($(X) + (105:2.04cm) $) ($(X) + (120:2.04cm) $) ($(X) + (135:2.04cm) $) ($(X) + (150:2.04cm) $) ($(X) + (165:2.04cm) $) ($(X) + (180:2.04cm) $) ($(X) + (195:2.04cm) $) ($(X) + (210:2.04cm) $) ($(X) + (225:2.04cm) $) ($(X) + (240:2.04cm) $) ($(X) + (255:2.04cm) $) ($(X) + (270:2.04cm) $) ($(X) + (285:2.04cm) $) ($(X) + (300:2.04cm) $) ($(X) + (315:2.04cm) $) ($(X) + (330:2.04cm) $) ($(X) + (345:1.56cm) $) };
\draw[opacity = 0.6] plot [smooth cycle, tension = 0.5] coordinates{($(X) + (0:1.3cm) $) ($(X) + (15:1.4cm) $) ($(X) + (30:1.3cm) $) ($(X) + (45:1.25cm) $) ($(X) + (60:1.35cm) $) ($(X) + (75:1.75cm) $) ($(X) + (90:1.8cm) $) ($(X) + (105:1.8cm) $) ($(X) + (120:1.8cm) $) ($(X) + (135:1.8cm) $) ($(X) + (150:1.8cm) $) ($(X) + (165:1.8cm) $) ($(X) + (180:1.8cm) $) ($(X) + (195:1.8cm) $) ($(X) + (210:1.8cm) $) ($(X) + (225:1.8cm) $) ($(X) + (240:1.8cm) $) ($(X) + (255:1.8cm) $) ($(X) + (270:1.8cm) $) ($(X) + (285:1.8cm) $) ($(X) + (300:1.8cm) $) ($(X) + (315:1.8cm) $) ($(X) + (330:1.8cm) $) ($(X) + (345:1.4cm) $) };
\draw[opacity = 0.6] plot [smooth cycle, tension = 0.5] coordinates{($(X) + (0:1.16cm) $) ($(X) + (15:1.24cm) $) ($(X) + (30:1.16cm) $) ($(X) + (45:1.12cm) $) ($(X) + (60:1.2cm) $) ($(X) + (75:1.52cm) $) ($(X) + (90:1.56cm) $) ($(X) + (105:1.56cm) $) ($(X) + (120:1.56cm) $) ($(X) + (135:1.56cm) $) ($(X) + (150:1.56cm) $) ($(X) + (165:1.56cm) $) ($(X) + (180:1.56cm) $) ($(X) + (195:1.56cm) $) ($(X) + (210:1.56cm) $) ($(X) + (225:1.56cm) $) ($(X) + (240:1.56cm) $) ($(X) + (255:1.56cm) $) ($(X) + (270:1.56cm) $) ($(X) + (285:1.56cm) $) ($(X) + (300:1.56cm) $) ($(X) + (315:1.56cm) $) ($(X) + (330:1.56cm) $) ($(X) + (345:1.24cm) $) };
\draw[opacity = 0.6] plot [smooth cycle, tension = 0.5] coordinates{($(X) + (0:1.02cm) $) ($(X) + (15:1.08cm) $) ($(X) + (30:1.02cm) $) ($(X) + (45:0.99cm) $) ($(X) + (60:1.05cm) $) ($(X) + (75:1.29cm) $) ($(X) + (90:1.32cm) $) ($(X) + (105:1.32cm) $) ($(X) + (120:1.32cm) $) ($(X) + (135:1.32cm) $) ($(X) + (150:1.32cm) $) ($(X) + (165:1.32cm) $) ($(X) + (180:1.32cm) $) ($(X) + (195:1.32cm) $) ($(X) + (210:1.32cm) $) ($(X) + (225:1.32cm) $) ($(X) + (240:1.32cm) $) ($(X) + (255:1.32cm) $) ($(X) + (270:1.32cm) $) ($(X) + (285:1.32cm) $) ($(X) + (300:1.32cm) $) ($(X) + (315:1.32cm) $) ($(X) + (330:1.32cm) $) ($(X) + (345:1.08cm) $) };
\draw[opacity = 0.6] plot [smooth cycle, tension = 0.5] coordinates{($(X) + (0:0.88cm) $) ($(X) + (15:0.92cm) $) ($(X) + (30:0.88cm) $) ($(X) + (45:0.86cm) $) ($(X) + (60:0.9cm) $) ($(X) + (75:1.06cm) $) ($(X) + (90:1.08cm) $) ($(X) + (105:1.08cm) $) ($(X) + (120:1.08cm) $) ($(X) + (135:1.08cm) $) ($(X) + (150:1.08cm) $) ($(X) + (165:1.08cm) $) ($(X) + (180:1.08cm) $) ($(X) + (195:1.08cm) $) ($(X) + (210:1.08cm) $) ($(X) + (225:1.08cm) $) ($(X) + (240:1.08cm) $) ($(X) + (255:1.08cm) $) ($(X) + (270:1.08cm) $) ($(X) + (285:1.08cm) $) ($(X) + (300:1.08cm) $) ($(X) + (315:1.08cm) $) ($(X) + (330:1.08cm) $) ($(X) + (345:0.92cm) $) };
\draw[opacity = 0.6] plot [smooth cycle, tension = 0.5] coordinates{($(X) + (0:0.74cm) $) ($(X) + (15:0.76cm) $) ($(X) + (30:0.74cm) $) ($(X) + (45:0.73cm) $) ($(X) + (60:0.75cm) $) ($(X) + (75:0.83cm) $) ($(X) + (90:0.84cm) $) ($(X) + (105:0.84cm) $) ($(X) + (120:0.84cm) $) ($(X) + (135:0.84cm) $) ($(X) + (150:0.84cm) $) ($(X) + (165:0.84cm) $) ($(X) + (180:0.84cm) $) ($(X) + (195:0.84cm) $) ($(X) + (210:0.84cm) $) ($(X) + (225:0.84cm) $) ($(X) + (240:0.84cm) $) ($(X) + (255:0.84cm) $) ($(X) + (270:0.84cm) $) ($(X) + (285:0.84cm) $) ($(X) + (300:0.84cm) $) ($(X) + (315:0.84cm) $) ($(X) + (330:0.84cm) $) ($(X) + (345:0.76cm) $) };
\draw[opacity = 0.6] plot [smooth cycle, tension = 0.5] coordinates{($(X) + (0:0.6cm) $) ($(X) + (15:0.6cm) $) ($(X) + (30:0.6cm) $) ($(X) + (45:0.6cm) $) ($(X) + (60:0.6cm) $) ($(X) + (75:0.6cm) $) ($(X) + (90:0.6cm) $) ($(X) + (105:0.6cm) $) ($(X) + (120:0.6cm) $) ($(X) + (135:0.6cm) $) ($(X) + (150:0.6cm) $) ($(X) + (165:0.6cm) $) ($(X) + (180:0.6cm) $) ($(X) + (195:0.6cm) $) ($(X) + (210:0.6cm) $) ($(X) + (225:0.6cm) $) ($(X) + (240:0.6cm) $) ($(X) + (255:0.6cm) $) ($(X) + (270:0.6cm) $) ($(X) + (285:0.6cm) $) ($(X) + (300:0.6cm) $) ($(X) + (315:0.6cm) $) ($(X) + (330:0.6cm) $) ($(X) + (345:0.6cm) $) };

% the contours around the point (Y)

\draw[opacity = 0.6] plot [smooth cycle, tension = 0.6] coordinates{($(Y) + (0:1.74cm) $) ($(Y) + (15:1.74cm) $) ($(Y) + (30:1.74cm) $) ($(Y) + (45:1.74cm) $) ($(Y) + (60:1.74cm) $) ($(Y) + (75:1.74cm) $) ($(Y) + (90:1.74cm) $) ($(Y) + (105:1.74cm) $) ($(Y) + (120:1.4cm) $) ($(Y) + (135:1.31cm) $) ($(Y) + (150:1.49cm) $) ($(Y) + (165:1.7cm) $) ($(Y) + (180:1.7cm) $) ($(Y) + (195:1.7cm) $) ($(Y) + (210:1.7cm) $) ($(Y) + (225:1.74cm) $) ($(Y) + (240:1.74cm) $) ($(Y) + (255:1.74cm) $) ($(Y) + (270:1.74cm) $) ($(Y) + (285:1.74cm) $) ($(Y) + (300:1.74cm) $) ($(Y) + (315:1.74cm) $) ($(Y) + (330:1.74cm) $) ($(Y) + (345:1.74cm) $) };
\draw[opacity = 0.6] plot [smooth cycle, tension = 0.6] coordinates{($(Y) + (0:1.49cm) $) ($(Y) + (15:1.49cm) $) ($(Y) + (30:1.49cm) $) ($(Y) + (45:1.49cm) $) ($(Y) + (60:1.49cm) $) ($(Y) + (75:1.49cm) $) ($(Y) + (90:1.49cm) $) ($(Y) + (105:1.49cm) $) ($(Y) + (120:1.2cm) $) ($(Y) + (135:1.13cm) $) ($(Y) + (150:1.27cm) $) ($(Y) + (165:1.45cm) $) ($(Y) + (180:1.45cm) $) ($(Y) + (195:1.45cm) $) ($(Y) + (210:1.45cm) $) ($(Y) + (225:1.49cm) $) ($(Y) + (240:1.49cm) $) ($(Y) + (255:1.49cm) $) ($(Y) + (270:1.49cm) $) ($(Y) + (285:1.49cm) $) ($(Y) + (300:1.49cm) $) ($(Y) + (315:1.49cm) $) ($(Y) + (330:1.49cm) $) ($(Y) + (345:1.49cm) $) };
\draw[opacity = 0.6] plot [smooth cycle, tension = 0.6] coordinates{($(Y) + (0:1.23cm) $) ($(Y) + (15:1.23cm) $) ($(Y) + (30:1.23cm) $) ($(Y) + (45:1.23cm) $) ($(Y) + (60:1.23cm) $) ($(Y) + (75:1.23cm) $) ($(Y) + (90:1.23cm) $) ($(Y) + (105:1.23cm) $) ($(Y) + (120:1.0cm) $) ($(Y) + (135:0.94cm) $) ($(Y) + (150:1.06cm) $) ($(Y) + (165:1.2cm) $) ($(Y) + (180:1.2cm) $) ($(Y) + (195:1.2cm) $) ($(Y) + (210:1.2cm) $) ($(Y) + (225:1.23cm) $) ($(Y) + (240:1.23cm) $) ($(Y) + (255:1.23cm) $) ($(Y) + (270:1.23cm) $) ($(Y) + (285:1.23cm) $) ($(Y) + (300:1.23cm) $) ($(Y) + (315:1.23cm) $) ($(Y) + (330:1.23cm) $) ($(Y) + (345:1.23cm) $) };
\draw[opacity = 0.6] plot [smooth cycle, tension = 0.6] coordinates{($(Y) + (0:0.97cm) $) ($(Y) + (15:0.97cm) $) ($(Y) + (30:0.97cm) $) ($(Y) + (45:0.97cm) $) ($(Y) + (60:0.97cm) $) ($(Y) + (75:0.97cm) $) ($(Y) + (90:0.97cm) $) ($(Y) + (105:0.97cm) $) ($(Y) + (120:0.8cm) $) ($(Y) + (135:0.76cm) $) ($(Y) + (150:0.84cm) $) ($(Y) + (165:0.95cm) $) ($(Y) + (180:0.95cm) $) ($(Y) + (195:0.95cm) $) ($(Y) + (210:0.95cm) $) ($(Y) + (225:0.97cm) $) ($(Y) + (240:0.97cm) $) ($(Y) + (255:0.97cm) $) ($(Y) + (270:0.97cm) $) ($(Y) + (285:0.97cm) $) ($(Y) + (300:0.97cm) $) ($(Y) + (315:0.97cm) $) ($(Y) + (330:0.97cm) $) ($(Y) + (345:0.97cm) $) };
\draw[opacity = 0.6] plot [smooth cycle, tension = 0.6] coordinates{($(Y) + (0:0.71cm) $) ($(Y) + (15:0.71cm) $) ($(Y) + (30:0.71cm) $) ($(Y) + (45:0.71cm) $) ($(Y) + (60:0.71cm) $) ($(Y) + (75:0.71cm) $) ($(Y) + (90:0.71cm) $) ($(Y) + (105:0.71cm) $) ($(Y) + (120:0.6cm) $) ($(Y) + (135:0.57cm) $) ($(Y) + (150:0.63cm) $) ($(Y) + (165:0.7cm) $) ($(Y) + (180:0.7cm) $) ($(Y) + (195:0.7cm) $) ($(Y) + (210:0.7cm) $) ($(Y) + (225:0.71cm) $) ($(Y) + (240:0.71cm) $) ($(Y) + (255:0.71cm) $) ($(Y) + (270:0.71cm) $) ($(Y) + (285:0.71cm) $) ($(Y) + (300:0.71cm) $) ($(Y) + (315:0.71cm) $) ($(Y) + (330:0.71cm) $) ($(Y) + (345:0.71cm) $) };
\draw[opacity = 0.6] plot [smooth cycle, tension = 0.6] coordinates{($(Y) + (0:0.46cm) $) ($(Y) + (15:0.46cm) $) ($(Y) + (30:0.46cm) $) ($(Y) + (45:0.46cm) $) ($(Y) + (60:0.46cm) $) ($(Y) + (75:0.46cm) $) ($(Y) + (90:0.46cm) $) ($(Y) + (105:0.46cm) $) ($(Y) + (120:0.4cm) $) ($(Y) + (135:0.39cm) $) ($(Y) + (150:0.41cm) $) ($(Y) + (165:0.45cm) $) ($(Y) + (180:0.45cm) $) ($(Y) + (195:0.45cm) $) ($(Y) + (210:0.45cm) $) ($(Y) + (225:0.46cm) $) ($(Y) + (240:0.46cm) $) ($(Y) + (255:0.46cm) $) ($(Y) + (270:0.46cm) $) ($(Y) + (285:0.46cm) $) ($(Y) + (300:0.46cm) $) ($(Y) + (315:0.46cm) $) ($(Y) + (330:0.46cm) $) ($(Y) + (345:0.46cm) $) };
\draw[opacity = 0.6] plot [smooth cycle, tension = 0.6] coordinates{($(Y) + (0:0.2cm) $) ($(Y) + (15:0.2cm) $) ($(Y) + (30:0.2cm) $) ($(Y) + (45:0.2cm) $) ($(Y) + (60:0.2cm) $) ($(Y) + (75:0.2cm) $) ($(Y) + (90:0.2cm) $) ($(Y) + (105:0.2cm) $) ($(Y) + (120:0.2cm) $) ($(Y) + (135:0.2cm) $) ($(Y) + (150:0.2cm) $) ($(Y) + (165:0.2cm) $) ($(Y) + (180:0.2cm) $) ($(Y) + (195:0.2cm) $) ($(Y) + (210:0.2cm) $) ($(Y) + (225:0.2cm) $) ($(Y) + (240:0.2cm) $) ($(Y) + (255:0.2cm) $) ($(Y) + (270:0.2cm) $) ($(Y) + (285:0.2cm) $) ($(Y) + (300:0.2cm) $) ($(Y) + (315:0.2cm) $) ($(Y) + (330:0.2cm) $) ($(Y) + (345:0.2cm) $) };

%and now for the contours around (Z)

\draw[opacity = 0.6] plot [smooth cycle, tension = 0.6] coordinates{($(Z) + (0:1.23cm) $) ($(Z) + (15:1.23cm) $) ($(Z) + (30:1.23cm) $) ($(Z) + (45:1.23cm) $) ($(Z) + (60:1.23cm) $) ($(Z) + (75:1.23cm) $) ($(Z) + (90:1.23cm) $) ($(Z) + (105:1.23cm) $) ($(Z) + (120:1.23cm) $) ($(Z) + (135:1.23cm) $) ($(Z) + (150:1.11cm) $) ($(Z) + (165:0.95cm) $) ($(Z) + (180:0.83cm) $) ($(Z) + (195:0.76cm) $) ($(Z) + (210:0.73cm) $) ($(Z) + (225:0.73cm) $) ($(Z) + (240:0.75cm) $) ($(Z) + (255:0.83cm) $) ($(Z) + (270:0.95cm) $) ($(Z) + (285:1.03cm) $) ($(Z) + (300:1.03cm) $) ($(Z) + (315:1.07cm) $) ($(Z) + (330:1.11cm) $) ($(Z) + (345:1.15cm) $) };
\draw[opacity = 0.6] plot [smooth cycle, tension = 0.6] coordinates{($(Z) + (0:0.96cm) $) ($(Z) + (15:0.96cm) $) ($(Z) + (30:0.96cm) $) ($(Z) + (45:0.96cm) $) ($(Z) + (60:0.96cm) $) ($(Z) + (75:0.96cm) $) ($(Z) + (90:0.96cm) $) ($(Z) + (105:0.96cm) $) ($(Z) + (120:0.96cm) $) ($(Z) + (135:0.96cm) $) ($(Z) + (150:0.87cm) $) ($(Z) + (165:0.75cm) $) ($(Z) + (180:0.66cm) $) ($(Z) + (195:0.61cm) $) ($(Z) + (210:0.58cm) $) ($(Z) + (225:0.59cm) $) ($(Z) + (240:0.6cm) $) ($(Z) + (255:0.66cm) $) ($(Z) + (270:0.75cm) $) ($(Z) + (285:0.81cm) $) ($(Z) + (300:0.81cm) $) ($(Z) + (315:0.84cm) $) ($(Z) + (330:0.87cm) $) ($(Z) + (345:0.9cm) $) };
\draw[opacity = 0.6] plot [smooth cycle, tension = 0.6] coordinates{($(Z) + (0:0.69cm) $) ($(Z) + (15:0.69cm) $) ($(Z) + (30:0.69cm) $) ($(Z) + (45:0.69cm) $) ($(Z) + (60:0.69cm) $) ($(Z) + (75:0.69cm) $) ($(Z) + (90:0.69cm) $) ($(Z) + (105:0.69cm) $) ($(Z) + (120:0.69cm) $) ($(Z) + (135:0.69cm) $) ($(Z) + (150:0.63cm) $) ($(Z) + (165:0.55cm) $) ($(Z) + (180:0.49cm) $) ($(Z) + (195:0.45cm) $) ($(Z) + (210:0.44cm) $) ($(Z) + (225:0.44cm) $) ($(Z) + (240:0.45cm) $) ($(Z) + (255:0.49cm) $) ($(Z) + (270:0.55cm) $) ($(Z) + (285:0.59cm) $) ($(Z) + (300:0.59cm) $) ($(Z) + (315:0.61cm) $) ($(Z) + (330:0.63cm) $) ($(Z) + (345:0.65cm) $) };
\draw[opacity = 0.6] plot [smooth cycle, tension = 0.6] coordinates{($(Z) + (0:0.42cm) $) ($(Z) + (15:0.42cm) $) ($(Z) + (30:0.42cm) $) ($(Z) + (45:0.42cm) $) ($(Z) + (60:0.42cm) $) ($(Z) + (75:0.42cm) $) ($(Z) + (90:0.42cm) $) ($(Z) + (105:0.42cm) $) ($(Z) + (120:0.42cm) $) ($(Z) + (135:0.42cm) $) ($(Z) + (150:0.39cm) $) ($(Z) + (165:0.35cm) $) ($(Z) + (180:0.32cm) $) ($(Z) + (195:0.3cm) $) ($(Z) + (210:0.29cm) $) ($(Z) + (225:0.3cm) $) ($(Z) + (240:0.3cm) $) ($(Z) + (255:0.32cm) $) ($(Z) + (270:0.35cm) $) ($(Z) + (285:0.37cm) $) ($(Z) + (300:0.37cm) $) ($(Z) + (315:0.38cm) $) ($(Z) + (330:0.39cm) $) ($(Z) + (345:0.4cm) $) };
\draw[opacity = 0.6] plot [smooth cycle, tension = 0.6] coordinates{($(Z) + (0:0.15cm) $) ($(Z) + (15:0.15cm) $) ($(Z) + (30:0.15cm) $) ($(Z) + (45:0.15cm) $) ($(Z) + (60:0.15cm) $) ($(Z) + (75:0.15cm) $) ($(Z) + (90:0.15cm) $) ($(Z) + (105:0.15cm) $) ($(Z) + (120:0.15cm) $) ($(Z) + (135:0.15cm) $) ($(Z) + (150:0.15cm) $) ($(Z) + (165:0.15cm) $) ($(Z) + (180:0.15cm) $) ($(Z) + (195:0.15cm) $) ($(Z) + (210:0.15cm) $) ($(Z) + (225:0.15cm) $) ($(Z) + (240:0.15cm) $) ($(Z) + (255:0.15cm) $) ($(Z) + (270:0.15cm) $) ($(Z) + (285:0.15cm) $) ($(Z) + (300:0.15cm) $) ($(Z) + (315:0.15cm) $) ($(Z) + (330:0.15cm) $) ($(Z) + (345:0.15cm) $) };

% finally, the outer contours

\draw[opacity = 0.6] plot [smooth cycle, tension = 0.5] coordinates{($(W) + (0:4.61cm) $) ($(W) + (15:4.11cm) $) ($(W) + (30:3.39cm) $) ($(W) + (45:3.12cm) $) ($(W) + (60:3.3cm) $) ($(W) + (75:3.37cm) $) ($(W) + (90:3.21cm) $) ($(W) + (105:2.94cm) $) ($(W) + (120:3.21cm) $) ($(W) + (135:3.66cm) $) ($(W) + (150:4.02cm) $) ($(W) + (165:4.38cm) $) ($(W) + (180:4.61cm) $) ($(W) + (195:4.7cm) $) ($(W) + (210:4.7cm) $) ($(W) + (225:4.51cm) $) ($(W) + (240:4.2cm) $) ($(W) + (255:3.79cm) $) ($(W) + (270:3.41cm) $) ($(W) + (285:2.98cm) $) ($(W) + (300:2.94cm) $) ($(W) + (315:3.79cm) $) ($(W) + (330:4.42cm) $) ($(W) + (345:4.7cm) $) };
\draw[opacity = 0.6] plot [smooth cycle, tension = 0.5] coordinates{($(W) + (0:4.76cm) $) ($(W) + (15:4.32cm) $) ($(W) + (30:3.68cm) $) ($(W) + (45:3.44cm) $) ($(W) + (60:3.6cm) $) ($(W) + (75:3.66cm) $) ($(W) + (90:3.52cm) $) ($(W) + (105:3.28cm) $) ($(W) + (120:3.52cm) $) ($(W) + (135:3.92cm) $) ($(W) + (150:4.24cm) $) ($(W) + (165:4.56cm) $) ($(W) + (180:4.76cm) $) ($(W) + (195:4.84cm) $) ($(W) + (210:4.84cm) $) ($(W) + (225:4.68cm) $) ($(W) + (240:4.4cm) $) ($(W) + (255:4.04cm) $) ($(W) + (270:3.7cm) $) ($(W) + (285:3.32cm) $) ($(W) + (300:3.28cm) $) ($(W) + (315:4.04cm) $) ($(W) + (330:4.6cm) $) ($(W) + (345:4.84cm) $) };
\draw[opacity = 0.6] plot [smooth cycle, tension = 0.5] coordinates{($(W) + (0:4.92cm) $) ($(W) + (15:4.53cm) $) ($(W) + (30:3.97cm) $) ($(W) + (45:3.76cm) $) ($(W) + (60:3.9cm) $) ($(W) + (75:3.96cm) $) ($(W) + (90:3.83cm) $) ($(W) + (105:3.62cm) $) ($(W) + (120:3.83cm) $) ($(W) + (135:4.18cm) $) ($(W) + (150:4.46cm) $) ($(W) + (165:4.74cm) $) ($(W) + (180:4.92cm) $) ($(W) + (195:4.98cm) $) ($(W) + (210:4.98cm) $) ($(W) + (225:4.84cm) $) ($(W) + (240:4.6cm) $) ($(W) + (255:4.29cm) $) ($(W) + (270:3.98cm) $) ($(W) + (285:3.66cm) $) ($(W) + (300:3.62cm) $) ($(W) + (315:4.29cm) $) ($(W) + (330:4.78cm) $) ($(W) + (345:4.98cm) $) };
\draw[opacity = 0.6] plot [smooth cycle, tension = 0.5] coordinates{($(W) + (0:5.07cm) $) ($(W) + (15:4.74cm) $) ($(W) + (30:4.26cm) $) ($(W) + (45:4.08cm) $) ($(W) + (60:4.2cm) $) ($(W) + (75:4.25cm) $) ($(W) + (90:4.14cm) $) ($(W) + (105:3.96cm) $) ($(W) + (120:4.14cm) $) ($(W) + (135:4.44cm) $) ($(W) + (150:4.68cm) $) ($(W) + (165:4.92cm) $) ($(W) + (180:5.07cm) $) ($(W) + (195:5.13cm) $) ($(W) + (210:5.13cm) $) ($(W) + (225:5.01cm) $) ($(W) + (240:4.8cm) $) ($(W) + (255:4.53cm) $) ($(W) + (270:4.27cm) $) ($(W) + (285:3.99cm) $) ($(W) + (300:3.96cm) $) ($(W) + (315:4.53cm) $) ($(W) + (330:4.95cm) $) ($(W) + (345:5.13cm) $) };
\draw[opacity = 0.6] plot [smooth cycle, tension = 0.5] coordinates{($(W) + (0:5.22cm) $) ($(W) + (15:4.95cm) $) ($(W) + (30:4.55cm) $) ($(W) + (45:4.4cm) $) ($(W) + (60:4.5cm) $) ($(W) + (75:4.54cm) $) ($(W) + (90:4.45cm) $) ($(W) + (105:4.3cm) $) ($(W) + (120:4.45cm) $) ($(W) + (135:4.7cm) $) ($(W) + (150:4.9cm) $) ($(W) + (165:5.1cm) $) ($(W) + (180:5.22cm) $) ($(W) + (195:5.28cm) $) ($(W) + (210:5.28cm) $) ($(W) + (225:5.17cm) $) ($(W) + (240:5.0cm) $) ($(W) + (255:4.78cm) $) ($(W) + (270:4.56cm) $) ($(W) + (285:4.33cm) $) ($(W) + (300:4.3cm) $) ($(W) + (315:4.78cm) $) ($(W) + (330:5.13cm) $) ($(W) + (345:5.28cm) $) };
\draw[opacity = 0.6] plot [smooth cycle, tension = 0.5] coordinates{($(W) + (0:5.38cm) $) ($(W) + (15:5.16cm) $) ($(W) + (30:4.84cm) $) ($(W) + (45:4.72cm) $) ($(W) + (60:4.8cm) $) ($(W) + (75:4.83cm) $) ($(W) + (90:4.76cm) $) ($(W) + (105:4.64cm) $) ($(W) + (120:4.76cm) $) ($(W) + (135:4.96cm) $) ($(W) + (150:5.12cm) $) ($(W) + (165:5.28cm) $) ($(W) + (180:5.38cm) $) ($(W) + (195:5.42cm) $) ($(W) + (210:5.42cm) $) ($(W) + (225:5.34cm) $) ($(W) + (240:5.2cm) $) ($(W) + (255:5.02cm) $) ($(W) + (270:4.85cm) $) ($(W) + (285:4.66cm) $) ($(W) + (300:4.64cm) $) ($(W) + (315:5.02cm) $) ($(W) + (330:5.3cm) $) ($(W) + (345:5.42cm) $) };
\draw[opacity = 0.6] plot [smooth cycle, tension = 0.5] coordinates{($(W) + (0:5.54cm) $) ($(W) + (15:5.37cm) $) ($(W) + (30:5.13cm) $) ($(W) + (45:5.04cm) $) ($(W) + (60:5.1cm) $) ($(W) + (75:5.12cm) $) ($(W) + (90:5.07cm) $) ($(W) + (105:4.98cm) $) ($(W) + (120:5.07cm) $) ($(W) + (135:5.22cm) $) ($(W) + (150:5.34cm) $) ($(W) + (165:5.46cm) $) ($(W) + (180:5.54cm) $) ($(W) + (195:5.56cm) $) ($(W) + (210:5.56cm) $) ($(W) + (225:5.5cm) $) ($(W) + (240:5.4cm) $) ($(W) + (255:5.27cm) $) ($(W) + (270:5.14cm) $) ($(W) + (285:4.99cm) $) ($(W) + (300:4.98cm) $) ($(W) + (315:5.27cm) $) ($(W) + (330:5.47cm) $) ($(W) + (345:5.56cm) $) };
\draw[opacity = 0.6] plot [smooth cycle, tension = 0.5] coordinates{($(W) + (0:5.69cm) $) ($(W) + (15:5.58cm) $) ($(W) + (30:5.42cm) $) ($(W) + (45:5.36cm) $) ($(W) + (60:5.4cm) $) ($(W) + (75:5.42cm) $) ($(W) + (90:5.38cm) $) ($(W) + (105:5.32cm) $) ($(W) + (120:5.38cm) $) ($(W) + (135:5.48cm) $) ($(W) + (150:5.56cm) $) ($(W) + (165:5.64cm) $) ($(W) + (180:5.69cm) $) ($(W) + (195:5.71cm) $) ($(W) + (210:5.71cm) $) ($(W) + (225:5.67cm) $) ($(W) + (240:5.6cm) $) ($(W) + (255:5.51cm) $) ($(W) + (270:5.42cm) $) ($(W) + (285:5.33cm) $) ($(W) + (300:5.32cm) $) ($(W) + (315:5.51cm) $) ($(W) + (330:5.65cm) $) ($(W) + (345:5.71cm) $) };
\draw[opacity = 0.6] plot [smooth cycle, tension = 0.5] coordinates{($(W) + (0:5.85cm) $) ($(W) + (15:5.79cm) $) ($(W) + (30:5.71cm) $) ($(W) + (45:5.68cm) $) ($(W) + (60:5.7cm) $) ($(W) + (75:5.71cm) $) ($(W) + (90:5.69cm) $) ($(W) + (105:5.66cm) $) ($(W) + (120:5.69cm) $) ($(W) + (135:5.74cm) $) ($(W) + (150:5.78cm) $) ($(W) + (165:5.82cm) $) ($(W) + (180:5.85cm) $) ($(W) + (195:5.86cm) $) ($(W) + (210:5.86cm) $) ($(W) + (225:5.83cm) $) ($(W) + (240:5.8cm) $) ($(W) + (255:5.75cm) $) ($(W) + (270:5.71cm) $) ($(W) + (285:5.67cm) $) ($(W) + (300:5.66cm) $) ($(W) + (315:5.75cm) $) ($(W) + (330:5.83cm) $) ($(W) + (345:5.86cm) $) };
\draw[opacity = 0.6] plot [smooth cycle, tension = 0.5] coordinates{($(W) + (0:6.0cm) $) ($(W) + (15:6.0cm) $) ($(W) + (30:6.0cm) $) ($(W) + (45:6.0cm) $) ($(W) + (60:6.0cm) $) ($(W) + (75:6.0cm) $) ($(W) + (90:6.0cm) $) ($(W) + (105:6.0cm) $) ($(W) + (120:6.0cm) $) ($(W) + (135:6.0cm) $) ($(W) + (150:6.0cm) $) ($(W) + (165:6.0cm) $) ($(W) + (180:6.0cm) $) ($(W) + (195:6.0cm) $) ($(W) + (210:6.0cm) $) ($(W) + (225:6.0cm) $) ($(W) + (240:6.0cm) $) ($(W) + (255:6.0cm) $) ($(W) + (270:6.0cm) $) ($(W) + (285:6.0cm) $) ($(W) + (300:6.0cm) $) ($(W) + (315:6.0cm) $) ($(W) + (330:6.0cm) $) ($(W) + (345:6.0cm) $) };

%insertion points and poles labels

\fill[white] ($(X) + (-0.4,0.4)$) circle (0.25);
\node at ($(X) + (-0.4,0.4)$) {\large $z_1$};
\fill[white] ($(Y) + (0.5,0.5)$) circle (0.25);
\node at ($(Y) + (0.5,0.5)$) {\large $z_2$};
\fill[white] ($(Z) + (0.5,0.5)$) circle (0.25);
\node at ($(Z) + (0.5,0.5)$) {\large $z_3$};

\fill[white] ($(A) + (-0.1,0.4)$) circle (0.25);
\node at ($(A) + (-0.1,0.4)$) {\large $a_1$};
\fill[white] ($(B) + (0.45,0.45)$) circle (0.25);
\node at ($(B) + (0.45,0.45)$) {\large $a_2$};
\fill[white] ($(C) + (-0.4,-0.1)$) circle (0.25);
\node at ($(C) + (-0.4,-0.1)$) {\large $a_3$};
\fill[white] ($(D) + (0.1,-0.6)$) circle (0.25);
\node at ($(D) + (0.1,-0.6)$) {\large $a_4$};

\end{tikzpicture}
\caption{The grey lines are the horizontal trajectories of a Strebel differential, see eq.~(\ref{hortraj}), while the coloured lines describe the critical horizontal trajectories that make up the critical Strebel graph, see Section~\ref{sec:6} for more details.  The (double) poles $z_i$ of the Strebel differential are denoted by black dots ($z_4=\infty$), while the zeros $a_i$ are represented by black crosses, see eq.~(\ref{y0square}). Finally, the solid black lines between the poles describe the dual edges to the critical Strebel graph, and therefore correspond to the edges of the skeleton graph of the field theory. }
\label{fig:1}
\end{figure}

Strebel differentials $\phi_S(z)dz^2$ are meromorphic quadratic differentials on a Riemann surface $\Sigma_{g,n}$ with double poles at the $n$ punctures. Moreover, the $(6g-6+3n)$ independent Strebel lengths between any two of the $(4g+2n-4)$ (generically distinct) zeroes are always real, 
\be\label{streb-length0}
\int_{a_k}^{a_m}\sqrt{\phi_S(z)} = l_{km} \in \mathbb{R_+} \ .
\ee
A theorem due to Strebel tells us that there is a unique such Strebel differential for every point on 
${\cal M}_{g,n}$ if one fixes the $n$ residues at the poles (which are necessarily real since they are linear combinations of the $l_{km}$). This is a bijective correspondence, i.e.\ if one specifies the 
$(6g-6+3n)$ Strebel lengths then one lands on a unique point in ${\cal M}_{g,n}\times {\mathbb R}_+^n$ and vice versa. Thus, if we fix the residues, the remaining $(6g-6+2n)$ independent real Strebel lengths $l_{km}$ can be viewed as a particular parametrisation of the string moduli space  ${\cal M}_{g,n}$. Furthermore, there is a so-called critical graph associated with the Strebel differential which has $n$ faces (with the topology of a disk), each of which contains exactly one double pole of the Strebel differential; in addition it has $(4g+2n-4)$ vertices which are the zeroes of the Strebel differential, and $(6g+3n-6)$ edges which connect the vertices, see Fig~\ref{fig:1}. The idea of \cite{Gopakumar:2005fx} was then that this critical graph is to be identified with the dual graph to the skeleton diagram of the field theory; the skeleton graph itself is then represented by the solid black lines in Fig~\ref{fig:1}. In fact, this construction can be seen as a way of implementing open-closed string duality, with the open string ribbon graphs of the field theory being glued along the critical Strebel graph to form the closed string worldsheet \cite{Gopakumar:2005fx}.

In \cite{Gopakumar:2005fx}, an additional identification was proposed between the (inverse) Schwinger proper times of the Feynman diagrams and the Strebel lengths (\ref{streb-length0}). This was a concrete way to promote the field theory answer into a dual world-sheet correlator; in particular it leads to a specific candidate integrand on moduli space --- the RHS of (\ref{corresp}). 
However, it was pointed out in 
\cite{Aharony:2006th} that this prescription had the disadvantage of not manifestly preserving the global spacetime special conformal symmetry of the putative world-sheet correlators. An alternative prescription was put forward by Razamat \cite{Razamat:2008zr} which associated the number $n_{ij}$ of homotopic Wick contractions between a pair of vertices, to the corresponding Strebel lengths (\ref{streb-length0}) of the dual edges. This discrete prescription was particularly well suited for the zero-dimensional Gaussian matrix model (where the correlators do not carry any spacetime dependence), and was explored in 
 \cite{Razamat:2008zr, Razamat:2009mc,Gopakumar:2011ev, Gopakumar:2012ny, Koch:2014hsa}.  

To summarise, the broad thrust of  \cite{Gopakumar:2003ns, Gopakumar:2004qb, Gopakumar:2005fx, 
Razamat:2008zr} was a  definite prescription\footnote{This prescription was generalised to theories with fundamental matter in \cite{Yaakov:2006ce} and also applied to a number of different correlators, see \cite{Aharony:2006th, Furuuchi:2005qm, David:2006qc, Aharony:2007fs, David:2008iz, Lal:2020xwc} for a partial list.}  by which the mechanism of open-closed string duality is realised. It associates to individual Feynman diagrams of the field theory, specific points in the moduli space of the dual closed string theory, thus giving a constructive method to go from the LHS to the RHS in (\ref{corresp}). 
\smallskip

In this paper we will check this proposal in an example of the ${\rm AdS}/{\rm CFT}$ correspondence which can serve as a concrete testbed for understanding the precise working of the duality and its dictionary. It has recently been understood that string theory on ${\rm AdS}_3\times {\rm S}^3\times \mathbb{T}^4$ with one unit ($k=1$) of NS-NS flux is dual to the free symmetric orbifold CFT ${\rm Sym}^K(\mathbb{T}^4)$ in the large $K$ limit. The full perturbative string spectrum exactly agrees with that of the symmetric orbifold \cite{Eberhardt:2018ouy} (earlier work on the $k=1$ theory includes  \cite{Gaberdiel:2018rqv,Giveon:2005mi,Giribet:2018ada,Gaberdiel:2017oqg,Ferreira:2017pgt}). Furthermore, it was shown  in \cite{Eberhardt:2019ywk} how correlators in the world-sheet string theory localise on moduli space to special points which admit a covering map of the ${\rm AdS}_3$ boundary ${\rm S}^2$, thereby manifestly reproducing the symmetric orbifold correlators that can be calculated using such a covering map approach \cite{Lunin:2000yv, Lunin:2001pw}. 
This was generalised beyond genus zero on the world-sheet in \cite{Eberhardt:2020akk}, and to geometries with other boundaries in \cite{Eberhardt:2020bgq}. Recently, in \cite{Dei:2020zui} the crucial property of world-sheet localisation was shown to follow from a twistorial incidence relation in a free field realisation of the $k=1$ world-sheet sigma model. In this example we therefore understand in detail how the RHS of (\ref{corresp}) gives rise to the LHS.

This example therefore allows us to analyse whether the reconstruction proposal of \cite{Gopakumar:2003ns, Gopakumar:2004qb, Gopakumar:2005fx} will indeed reproduce the correct world-sheet theory.\footnote{It would be interesting to see how to generalise the approach here to theories whose bulk duals are not string theories, see recent work on reconstructing the bulk and thus deriving the AdS/CFT correspondence for vector-like large $N$ models in \cite{deMelloKoch:2018ivk,oferetal} and references therein.}  
 To this end, we will start with an $n$-point twisted sector correlator in the orbifold CFT  --- i.e.\ the LHS of (\ref{corresp}) ---  
\be\label{symorb0}
\langle  \sigma_{w_1}(x_1) \cdots \sigma_{w_n}(x_n) \rangle \ ,
\ee
and try to rewrite it as a world-sheet integral. 
The key idea is to study (\ref{symorb0}) in a special Gross-Mende like limit \cite{Gross:1987ar}, where the twist\footnote{This is not to be confused with $\tau=\Delta-J$. In our context `twist' refers to the twisted sectors of the symmetric orbifold.} labels $w_i$ of the operators are taken large (but small compared to $K$).
The nice feature of this limit is that the covering maps which contribute to the correlator 
actually become dense on the moduli space of the covering surface. If we concentrate on the case where the covering surface (which will be identified with the world-sheet) is of genus zero, the computation of correlators via the covering map \cite{Lunin:2000yv, Lunin:2001pw} simplifies significantly. In fact, we will be able to map the problem of finding the different covering maps to that of solving for the large $N$ limit\footnote{This $N$ should not be confused with the parameter of the 't~Hooft genus  expansion of the orbifold theory --- that was denoted above by $K$, the rank of the symmetric group, and will be taken to $\infty$ as we will always be at genus zero. $N$ is more like a charge, proportional to the large twists $w_i$, see eq.~(\ref{RHform}), which is separately taken to be large.} of a special class of matrix models with Penner-like logarithmic potentials. 

The inequivalent covering maps for an $n$-point correlator will turn out to be para\-me\-trised by a set of $(2n-6)$ real parameters. These parameters determine the so-called spectral curve of the matrix model at leading order in $\frac{1}{N}$. This defines a meromorphic differential $y_0(z) dz$ whose square has double poles at the points $z_i$ on the covering space which correspond to the pre-images of the branch points $x_i$ in (\ref{symorb0}), see eq.~(\ref{streb}) 
\be\label{y0square}
y_0^2(z) dz^2 =  \frac{ \alpha_n^2\, dz^2}{\prod_{i=1}^{n}(z-z_i)^2} \prod_{k=1}^{2n-4}(z-a_k) = - 4\pi^2\phi_S(z) \, dz^2\ .
\ee
The $(2n-6)$ real parameters that specify the solution also determine the zeroes and poles of this differential, and are nothing other than the period integrals of $y_0(z)dz$ along inequivalent homology cycles on the covering surface.

The upshot of this analysis is therefore, quite remarkably, that the solution (\ref{y0square}) to the matrix model which determines the covering maps is (minus) a Strebel differential $\phi_S(z) dz^2$ on the covering space --- it naturally comes baked into the problem!  In the process we will also see that there is a one-to-one mapping between these solutions of the matrix model, and the Feynman diagrams of the orbifold theory that can be associated to every admissible covering map \cite{Pakman:2009zz}; this is exactly as envisaged in the proposal of \cite{Gopakumar:2005fx, Gopakumar:2004qb}. The Strebel lengths (\ref{streb-length0}), which are proportional to the periods, take arbitrary real values which are equal to $\frac{n_{ij}}{2N}$ to leading order in $\frac{1}{N}$, where  $n_{ij}$ is the number of lines between vertices. This is similar to what was proposed in \cite{Razamat:2008zr}, although it will also be clear that such a simple relation will not hold when one takes into account subleading orders in $\frac{1}{N}$. Thus at large $N$ the sum over admissible covering maps goes over to an integral, as the set of allowed points on moduli space becomes dense in ${\cal M}_{0,n}$.  
 
Having seen how the moduli space  of world-sheets on the RHS of (\ref{corresp}) emerges naturally from the different Feynman diagram contributions, one can then address the second aspect of the program of \cite{Gopakumar:2005fx}. Namely, to determine the integrand on moduli space on the RHS of (\ref{corresp})  from the field theory. This can, again, be read off from the symmetric orbifold calculation of Lunin-Mathur, see eq.~(\ref{corrcov}). In the large $N$ limit, the dominiant contribution turns out to be the Liouville action (\ref{liouv0}), and this weighting factor has  multiple fascinating interpretations. On the one hand, it is proportional to 
a Nambu-Goto action for the Strebel metric --- which is the natural metric on the world-sheet one can associate to the quadratic differential $\phi_S(z)dz^2$, see eqs.~(\ref{Liouv}) and (\ref{strebmet}). Another striking form for this action arises from the observation that, to leading order in $\frac{1}{N}$, the Strebel differential in (\ref{y0square}) is nothing other than the Schwarzian of the covering map. The action is therefore also proportional to the absolute value of the Schwarzian, see eq.~(\ref{Liouv2}), and this leads to a suggestive ${\rm AdS}_3$ generalisation of the universal ${\rm AdS}_2$ expression for the spacetime action \cite{Maldacena:2016upp}.  Finally, there is a third interpretation of the action from the spacetime ${\rm S}^2$ point of view which is very reminiscent of the action for a rigid string that had been proposed in pre-holographic days to describe large $N$ gauge theories \cite{Polyakov:1986cs, Kleinert:1986bk, Horava:1993aq},  see eq.~(\ref{rigid}).  

In \cite{Eberhardt:2019ywk} the above Liouville action on the symmetric orbifold was shown to agree with the classical action of the ${\rm AdS}_3$ string theory, evaluated on the covering maps onto which the world-sheet path integral localises. This therefore essentially completes the circle by showing how the broad procedure of \cite{Gopakumar:2004qb, Gopakumar:2005fx, Razamat:2008zr} applied to free field orbifold CFT correlators can reconstruct not only the string moduli space in a precise way, but also what the world-sheet correlators are, at least in the large twist limit. It is a proof of concept that it is possible to go in both directions between the two sides of (\ref{corresp}).  

\medskip

The paper is organised as follows. After a brief recap of the covering space approach to computing symmetric orbifold correlators in Section~\ref{sec:symm-corr}, we set up the problem of finding the relevant covering maps in the large twist limit in Section~\ref{sec:lrg-twist}. Section~ \ref{sec:MM} is the technical heart of the paper where the equations determining the covering map are mapped onto the saddle point equations of a Penner-like matrix model. We use conventional matrix model technology to find the saddle point solutions, but then interpret the parameters in a slightly different way than for usual matrix models. We also make some comments on the finite $N$ generalisation in Section~\ref{sec:4.3}. With these results in hand we are then ready to understand their consequences: in Section~\ref{sec:spec-streb} we use the connection between covering maps and Feynman diagrams to see that the free parameters that enter into the matrix model solution are proportional to the number of Wick contractions, to leading order in $\frac{1}{N}$. Further, in Section~\ref{sec:6} we see that (the square of) the spectral curve of the matrix model is (minus) the Strebel differential, again to leading order in $\frac{1}{N}$, and hence that the sum over covering maps goes over to an integral over the string moduli space. We comment on the relation of the Strebel differential to the Schwarzian, at finite $N$, in Section~\ref{sec:6.1}. Section~\ref{sec:worldsheet} then goes on to reconstruct the weights associated with each point in moduli space, and briefly discusses their different avatars. Finally, Section~\ref{sec:conclusions} contains a somewhat extended list of open questions, categorised by theme. There are a couple of appendices with some additional technical sidelights. 

\section{Correlators in the symmetric orbifold CFT}\label{sec:symm-corr}

In this section we give a brief review about how correlation functions of symmetric orbifold theories can be computed in terms of covering maps; this approach goes back to the work of Lunin \& Mathur \cite{Lunin:2000yv,Lunin:2001pw}. In the following we shall mainly concentrate on the twisted sector ground states, although the method is also believed to be applicable to descendant states, see eq.~(\ref{corrcov}) below. 

Let us start by recapitulating some basic features of symmetric orbifold theories. Let ${\cal S}$ be a conformal field theory --- the case we primarily have in mind is that ${\cal S}$ is the superconformal field theory associated to $\mathbb{T}^4$, but this will not matter for the following --- then the symmetric orbifold theory of ${\cal S}$ is obtained by considering the $K$-fold tensor product theory ${\cal S}^{\otimes K}$ and taking the orbifold by $S_K$, where $S_K$ permutes the $K$ copies. As is familiar from general orbifold constructions, the orbifold theory consists of the so-called untwisted sector --- these are the states in ${\cal S}^{\otimes K}$ that are invariant under the action of all generators in $S_K$ --- as well as twisted sectors. These twisted sectors are in one-to-one correspondence with the conjugacy classes of $S_K$. Every permutation in $S_K$ can be written in cycle decomposition, and the cycle shape is invariant under conjugation; thus the conjugacy classes of $S_K$ are in one-to-one correspondence with the cycle shapes, and thus with partitions of $K$. In the following we shall exclusively consider the `single cycle' conjugacy classes, i.e.\  those permutations that are conjugate to a single cyclic permutation, say of length $w$. Note that under the AdS/CFT correspondence the states from these single cycle twisted sectors correspond to single string states in AdS. We will always be working in the large $K$ limit so that there is no restriction on the range of $w$; this corresponds to the limit in which the string coupling constant $g_s$ of the dual string theory is small. 

Let $\sigma_w$ be the ground state of the single cycle twisted sector of length $w$, and let $\pi_w$ be a representative of the corresponding conjugacy class of permutations, i.e.\ $\pi_w = (j_1 \cdots j_{w})$, where $j_m\in \{1,\ldots,K\}$ and $j_m\neq j_n$ for $m\neq n$. The twisted sector state $\sigma_w$ has the property that  as we analytically continue a field $\phi_l$ from the $l$'th copy of ${\cal S}^{\otimes K}$ around $\sigma_w$, it gets transmuted into the field $\phi_{\pi_w(l)}$ associated to the $\pi_w(l)$'th copy, 
\be\label{twis}
V(\phi_l, e^{2\pi i} z) \, \sigma_w(0) = V(\phi_{\pi_w(l)}, z) \, \sigma_w(0) \ . 
\ee
The main quantity we want to calculate are correlation functions of such twisted sector ground states $\sigma_{w_i}$, 
\be\label{symorb}
\langle  \sigma_{w_1}(x_1) \cdots \sigma_{w_n}(x_n) \rangle \ ,
\ee
where $x_i$ are coordinates on the sphere ${\rm S}^2$. Such a correlation function is only non-zero if there exist representatives $\pi_{w_i}$ --- recall that the twisted sectors are associated to conjugacy classes, so there are many different $w$-cycle permutations $\pi_{w}= (j_1 \cdots j_{w})$ --- such that 
\be\label{permid}
\pi_{w_1} \cdots \pi_{w_n} = {\rm id} \ . 
\ee
Let us assume this to be the case, and let us take $\sigma_{w_i}$ to be the twisted sector field associated to this representative permutation $\pi_{w_i}$. (Note that for the calculation of the actual correlation function one also needs to include combinatorial factors, see e.g.\ \cite{Pakman:2009zz}.) 

As is familiar from general CFT considerations, it is often useful to consider not just the correlation function (\ref{symorb}) itself, but to insert in addition chiral fields $\phi_l$ from ${\cal S}^{\otimes K}$. (If ${\cal S}$ is the $\mathbb{T}^4$ theory, $\phi_l$ could, for example, be one of the $4$  free bosons associated to the $l$'th copy of ${\cal S}^{\otimes K}$.) It follows from (\ref{twis}) that with respect to these individual fields the correlation function is not single valued since the fields $\phi_l$ change identity as they go around the different twisted sector insertions. 

The basic idea of Lunin \& Mathur  \cite{Lunin:2000yv,Lunin:2001pw} is to use the conformal symmetry to lift these correlation functions via the holomorphic covering map $\Gamma: \Sigma \rightarrow {\rm S}^2$ to single-valued functions on the covering surface $\Sigma$. Here $\Sigma$ is another Riemann surface --- we shall mainly be interested in the situation where $\Sigma$ is also a sphere, although in general $\Sigma$ may have higher genus --- and the condition that the correlation functions become single valued on $\Sigma$ is that near $\Gamma^{-1}(x_i) = z_i$ we have 
\be\label{defprop}
\Gamma(z) = x_i + a_i^{\Gamma} (z-z_i)^{w_i} + \cdots   \quad \hbox{for $z\sim z_i$} \qquad (i=1,\ldots,n)\ . 
\ee
(Thus the covering surface looks like a $w_i$-fold multi-storey car park near $z_i$, and hence the fields $\phi_l$ become single valued near $z_i$.) We note that the genus $g$  of $\Sigma$ is determined by the Riemann-Hurwitz formula in terms of the degree $N$ of the covering map, i.e.\ the number of preimages of a generic point $x\in \mathds{CP}^1$, as
\begin{equation}\label{RHgenus}
g=1-N+\frac{1}{2}\sum_{j=1}^{n} (w_j-1) \ .
\end{equation}
 From the symmetric orbifold perspective $N$ is the number of active colors, i.e.\ the number of different $j_n$ appearing in all the permutations $\pi_{w_i}$ in eq.~(\ref{permid}).

Returning to the correlator, if the $\sigma_{w_i}$ are the twisted sector ground states, they disappear entirely from the discussion once one has gone to the covering surface, i.e.\ they are invisible to the chiral fields $\phi_l$. As a consequence, the resulting correlation function on the covering surface is the vacuum correlator, which is therefore equal to unity. Thus the covering map transforms the correlation function (\ref{symorb}) to a trivial amplitude, and (\ref{symorb}) is just equal to the conformal factor associated to this covering map transformation. This conformal factor can be calculated using the Liouville action
\be\label{liouv0}
S_{\rm L}[\Phi] = \frac{c}{48 \pi} \int d^2z \sqrt{g} \bigl( \, 2 \, \partial \Phi \, \bar\partial \Phi + R \, \Phi \Bigr) \ , 
\ee
where $c$ is the central charge of the `seed theory' ${\cal S}$, e.g.\ if ${\cal S}$ is the $\mathbb{T}^4$ or K3 theory $c=6$, and we have explicitly \cite{Lunin:2000yv,Lunin:2001pw}
\be\label{corrcov}
\langle  {\cal O}_{w_1}(x_1) \cdots  {\cal O}_{w_n}(x_n) \rangle = \sum_\Gamma W_{\Gamma} \prod_{i=1}^n |a_i^{\Gamma}|^{-2(h_i-h_i^0)}e^{-S_{\rm L}[{\Phi}_\Gamma]} \ , 
\ee
where
\be\label{phisol}
{{\Phi}}_\Gamma = \log \, \partial_z \Gamma(z)  + \log \, \partial_{\bar{z}} \bar{\Gamma}(\bar{z})  \ . 
\ee
Here we have written the expression for the more general case where ${\cal O}_{w_i}$ is some operator of conformal dimension $h_i$ in the $w_i$-twisted sector, i.e.\ not necessarily equal to the ground state $\sigma_{w_i}$ whose conformal dimension equals $h^0=c \frac{w^2-1}{24w}$, with  $c$ the central charge of ${\cal S}$. The parameters $a_i^{\Gamma}$ are the coefficients appearing in (\ref{defprop}), and the $W_{\Gamma}$ are expected to be constants independent of $\Gamma$ \cite{Dei:2020}. 
Here the sum is over all possible covering maps --- for a given choice of $w_i$ only a finite number of branched coverings exist. (In general the different covering maps will involve covering surfaces $\Sigma$ of different genus $g$, though.)

While this is, in principle, a very powerful method for the calculation of these twisted sector correlators, it requires finding the corresponding covering maps.  In general, this is a difficult problem, but as we are about to explain, it actually simplifies in the limit in which the $w_i$ become large.

\section{The large twist limit for branched covers}\label{sec:lrg-twist}

As we have seen in the previous section, the computation of correlators in the symmetric orbifold theory reduces to finding all the branched coverings of the $n$-punctured sphere ${\rm S}^2$ of the spacetime CFT.  We will restrict to the case where the genus in (\ref{RHgenus})  equals $g=0$, i.e.\ we will consider covering maps
\begin{equation}
\Gamma \;:\; \mathds{CP}^1 \longrightarrow \mathds{CP}^1 \ ,
\end{equation}
for which the degree $N$ is given by, see eq.~(\ref{RHgenus})  
\begin{equation}\label{RHform}
N =1+\frac{1}{2}\sum_{j=1}^{n} (w_j-1) \ .
\end{equation}
The number of covering maps up to equivalence, i.e.\ up to the composition with  M\"obius transformations, is finite; for example for $n=4$ and fixing the $x_i$ (but allowing for a M\"obius transformation to act on the $z_i$), it equals \cite{hurwitz}
\be\label{number}
\# (\hbox{branched coverings}) = {\rm Min}_j \, w_j(N+1-w_j) \ .
\ee
An explicit formula for the general case of $n>4$ does not seem to be known, but it follows from our analysis in Section~\ref{sec:spec-streb} below that the corresponding number scales as $N^{2n-6}$ for large $w_i$, see eq.~(\ref{numbergen}). The problem of explicitly determining these branched covers for given branching data $\{ z_i, w_i\}$ is difficult, even for a four-point function ($n=4$). We will show in this section that there is a systematic way in which to compute genus-zero covering maps for connected $n$-point functions of single-cycle twist fields, if we consider the limit of large twist for all the operators.\footnote{While there have been many analyses of correlators of heavy operators in ${\rm AdS}/{\rm CFT}$, a Gross-Mende like limit for extremal correaltors in 4d based on the prescription of \cite{Gopakumar:2005fx} was attempted in \cite{David:2008iz}. Recently a very interesting `large p' limit of half-BPS operators was taken with a view to study the analogue of the Gross-Mende regime(s) in ${\rm AdS}_5\times {\rm S}^5$ \cite{Aprile:2020luw}.} More specifically, we will always assume that the `rank' $K$ of the symmetric group $S_K$ is taken to infinity first, before we take the twists $w_i\leq K$ to infinity.\footnote{This will ensure that the states we are considering correspond to perturbative string states in the dual theory and not, for example, to D-branes which would backreact on the geometry.}

Before we continue let us pause to mention that this limit is the analogue, in the dual ${\rm AdS}_3$ space, of a Gross-Mende like limit \cite{Gross:1987ar} in that we are looking at the scattering of states with high (spacetime) conformal dimension or energy in ${\rm AdS}_3$. (This follows from the fact that  $h^0=c\, \frac{w^2-1}{24w}$ scales as $w$ in the large twist limit.) Our twisted sector ground state operators are the analogue of tachyonic or massless states in flat space. Since we are keeping the positions $x_i$ fixed, we are considering the analogue of fixed angle scattering. (This is to be contrasted with a Regge-type limit, which it would also be interesting to explore.) 

It follows from eq.~(\ref{number}) (or eq.~(\ref{numbergen}) for $n>4$) that we have a large number of covering maps in this limit, and one of our main results is that the sum over all of these covering maps becomes,  in a precise manner, an integral over the moduli space of the covering space (which is an $n$-punctured sphere with $n-3$ moduli). Furthermore, the weight with which these different covering maps contribute equals the Nambu-Goto action, with the world-sheet metric in Strebel gauge. Alternatively, one may write this action in terms of the modulus  of the Schwarzian of the covering map, see Section~\ref{sec:worldsheet} below. We expect this action to have a Gross-Mende saddle point as in flat space. \smallskip

Let us return to the problem at hand and determine the covering maps that contribute to the correlator (\ref{symorb}) in the large twist limit.  We shall assume that infinity is a generic point of the covering map, i.e.\ that $x_i \neq \infty$ for $i=1,\ldots, n$, and we shall denote by $\lambda_a$, $a=1\ldots, N$, the preimages of $\infty$, i.e.\ the poles of $\Gamma(z)$. In what follows, it will be convenient to use the 
M\"obius invariance on the covering space $\Sigma$ to fix $z_1=0$, $z_2=1$ and $z_n=\infty$.
Then the covering map takes the form of a rational function of degree $N$, 
\begin{equation}
\Gamma(z) =\frac{p_N(z)}{q_N(z)} = \frac{p_N(z)}{\prod_{a=1}^N(z-\lambda_a)} \ ,
\end{equation}
where both $p_N(z)$ and $q_N(z)$ are polynomials of degree $N$. We have chosen the latter, without loss of generality, to be a monic polynomial with $N$ distinct zeroes corresponding to the poles $\lambda_a$ of $\Gamma(z)$.  

Following \cite{Roumpedakis:2018tdb} we now observe that the poles and zeros of $\partial \Gamma(z)$ are determined as follows.\footnote{We should alert the reader that \cite{Roumpedakis:2018tdb} takes one of the branch points to be $x_n=\infty$, which implies a coalescence of the poles and the zeroes of 
$\partial \Gamma(z)$. As a result, the specific expressions in \cite{Roumpedakis:2018tdb} have a different power of $z$ in the denominator.} From (\ref{defprop}) it is clear that the only zeros of $\partial \Gamma(z)$ occur at $z=z_i$ with order $(w_i-1)$. On the other hand, the only poles appear at $z=\lambda_a$, and they are all double poles. Thus $\partial \Gamma(z)$ necessarily has the form 
\begin{equation}\label{partG}
\partial \Gamma(z) = M_{\Gamma} \, \frac{ \prod_{i=1}^{n-1}(z-z_i)^{w_i-1}}{\prod_{a=1}^{N}(z-\lambda_a)^{2}} \ ,
\end{equation}
where $M_{\Gamma}$ is a non-zero constant. Note that $\partial\Gamma$ then also has the right branching behaviour at infinity  since $\partial \Gamma(z) \sim z^{-w_n-1}$ as $z\to \infty$, see eq.~(\ref{RHform}).  It is furthermore clear that the residue of $\partial \Gamma(z)$ at $z=\lambda_a$, i.e.\ the coefficient of the simple poles $\frac{1}{z-\lambda_a}$, must vanish (since $\Gamma(z)$ does not contain a logarithmic term). This leads to the $N$ ``scattering equations" \cite{Roumpedakis:2018tdb}\footnote{This is, again, very similar to the Gross-Mende  equations of \cite{Gross:1987ar}.}
\begin{equation}\label{scatt}
\sum_{i=1}^{n-1} \frac{w_i-1}{\lambda_a-z_i}=\sum_{b\neq a}^{N} \frac{2}{\lambda_a-\lambda_b} \ , \qquad  (a=1,\ldots, N) \ . 
\end{equation}
On the face of it, this seems to give rise to $N$ equations for the $N$ unknowns $\lambda_a$, up to permutations. However, there are actually only $(N-1)$ independent equations here since the sum of all the residues vanishes necessarily. Thus we can solve, for example, for $\lambda_a$ for $a=1,\ldots,N-1$ in terms of $\lambda_N$. Once one has determined these $(N-1)$ $\lambda_a$,  one is left with three undetermined parameters for the covering map: $\lambda_N$, $M_{\Gamma}$ and the constant of integration in going from $\partial \Gamma(z)$ to $\Gamma(z)$. These are fixed, for instance, by requiring that $\Gamma(z_i) =x_i$ (for $i=1,2,n$), with $x_j \neq \infty$. 

Our key observation here is that in the large $N$ limit --- because of (\ref{RHform}) this is the limit  when we take the $w_i$ large ---  this equation has a natural matrix model interpretation: the $\lambda_a$ can be thought of as the eigenvalues of a matrix model, with the 
RHS being the usual eigenvalue repulsion while the LHS is playing the role of an external potential determined by the $z_i$. We will exploit this connection in what follows, working mainly at large $N$, but commenting on the finite $N$ case at various points.

\section{A Penner-like matrix model and its solution}\label{sec:MM}

In this section we use matrix model techniques in order to solve the eigenvalue problem (\ref{scatt}), which we rewrite as 
\begin{equation}\label{scatt2}
\frac{1}{2}\sum_{i=1}^{n-1}\frac{\alpha_i}{\lambda_a-z_i} = \frac{1}{N} \sum_{b\neq a}^{N} \frac{1}{\lambda_a-\lambda_b} \ ,  \qquad  \quad \Big[\alpha_i = \frac{w_i - 1}{N}\Big] \ ,
\end{equation}
where $\alpha_i$ is held fixed in the large twist limit. Note that (\ref{RHform}) implies that
\begin{equation}\label{sumalph}
\frac{1}{2}\sum_{i=1}^{n}\alpha_i= 1-\frac{1}{N} \quad \Longrightarrow \quad \sum_{i=1}^{n-1}\alpha_i =2 - \alpha_n -\frac{2}{N}\  .
\end{equation}
As noted in the previous section we actually have $N-1$ independent equations here leaving one additional undetermined parameter in the solution. We will see later, see the discussion around eq.~(\ref{schwtransf}),  how this works for the large $N$ solution. 

We want to interpret eq.~(\ref{scatt2}) as the  saddle-point equation for the large $N$ matrix integral 
\begin{equation}\label{matint}
\mathcal{Z} = \int [dM] e^{-N\,{\rm Tr}W(M)} = \int \prod_{a=1}^{N} \frac{d\lambda_a}{2\pi} \,\Delta^2(\{\lambda_b\}) \, e^{-N\sum_{a=1}^{N}W(\lambda_a)} \ , 
\end{equation}
where the potential has a logarithmic Penner-like form  
\begin{equation}\label{potential}
W(z)=\sum_{i=1}^{n-1}\alpha_i \log\,(z-z_i)  \ , 
\end{equation}
and therefore 
\begin{equation}\label{potderiv}
W'(z)=\sum_{i=1}^{n-1}\frac{\alpha_i }{(z-z_i)}  \ .
\end{equation}
Furthermore $\Delta(\lambda_a)$ is the usual Vandermonde determinant. 
The covering map is related to the matrix model as
\begin{equation}\label{coverlead}
\frac{1}{N}\log \left[M_{\Gamma}^{-1}\partial\Gamma(z)\right]= W(z) -\frac{2}{N} \sum_{a=1}^{N} \log (z-\lambda_a)  \ , 
\end{equation} 
where we have used (\ref{partG}).

We should remark that the matrix integral is, at least at this stage, just a convenient prop for solving the equations (\ref{scatt}). In particular, in our problem, the $\lambda_a$ (and $z_i$) are complex and so are not, strictly speaking, eigenvalues of a Hermitian matrix model.  What the mathematical correspondence to saddle point equations for matrix integrals suggests is a method to solve them in the large $N$ limit.  In fact, as we will indicate below, our question will dictate a slightly different angle and interpretation of the problem from the more traditional matrix model point of view. 

The class of matrix models with logarithmic potential as in (\ref{potential}) are known as Penner-like models and have appeared in a number of physics contexts, including recently in the connection between AGT and topological strings \cite{Dijkgraaf:2009pc, Itoyama:2009sc, Eguchi:2009gf, Schiappa:2009cc, Mironov:2009ib}.  
We will draw upon some of these results in what follows though, as mentioned, we will have a somewhat different interpretation. It will, nevertheless be interesting to see whether the appearance of the same matrix models, in the context of topological strings and Liouville theory in the above references as well as in the present context, is more than a technically fortuitous coincidence. 

In any case, we proceed by introducing a (normalised) density of ``eigenvalues"
\begin{equation}
\rho(\lambda) =\frac{1}{N}\sum_ {a=1}^N\delta(\lambda-\lambda_a)\ .
\end{equation}
We expect that in the large $N$ limit this will go over to a smooth function which has support over some set of curves (which we denote below by ${\cal C}$) on the complex plane. The equation (\ref{scatt2}) determining the covering maps becomes then (for $\lambda \in {\cal C}$) 
\begin{equation}\label{intgr-eq}
\frac{1}{2}W^{\prime}(\lambda) = P \int_{\cal C}\frac{\rho(\lambda^{\prime}) d\lambda^{\prime}}{\lambda-\lambda^{\prime}} \ , 
\end{equation}
where $P$ denotes the principal value. This integral equation for $\rho(\lambda)$, including its support, can now be solved using conventional matrix model technology. We will do so in the next subsection using the method of loop equations. The advantage of this method is that it generalises beyond the leading large $N$ limit. There is also an equivalent method for solving the  large $N$ equations in terms of a Riemann-Hilbert problem, which we outline in Appendix~\ref{app:RH}. 

Either way, we note that we expect to find a family of solutions corresponding to the large multiplicity of covering maps in the large twist limit, see eqs.~(\ref{number}) and (\ref{numbergen}). For any of these solutions for 
$\rho(\lambda)$, the corresponding covering map will be determined, at leading order in $N$, by (\ref{coverlead}) 
\begin{equation}\label{covmap}
\frac{1}{N}\log \left[M_{\Gamma}^{-1}\partial\Gamma(z)\right] = \sum_{i=1}^{n}\alpha_i \log\,(z-z_i) -2\int_{\cal C} d\lambda \, \rho (\lambda) \log (z-\lambda)\ .
\end{equation}
We will also see in Section~\ref{sec:4.3} that it will be natural to go somewhat beyond this leading large $N$ answer.

\subsection{Solving the Matrix Model}

We can convert the saddle point equations for the eigenvalue density into a set of equations for the resolvent --- which are known as loop equations  \cite{Wadia:1980rb} and can actually be written down for finite $N$. These are then solved in terms of an auxiliary spectral curve, see \cite{Marino:2004eq} for a very nice exposition of these techniques. For the multi-Penner potentials the corresponding solutions and spectral curves have been explicitly studied to leading order in $\frac{1}{N}$ in \cite{Schiappa:2009cc}. 

To obtain the spectral curve we first define the resolvent via 
\begin{equation}\label{wdef}
u(z) =\frac{1}{N} \sum_{a=1}^{N} \frac{1}{z-\lambda_a} = \int_{\cal C} \frac{\rho(\lambda) d\lambda}{z-\lambda} \ ,
\end{equation}
from which we can deduce the eigenvalue density (in the large $N$ limit) by the discontinuity across the support ${\cal C}$
\begin{equation}\label{evdens}
\rho(\lambda)=- \frac{1}{2\pi i} \Bigl[ u(\lambda+i\epsilon)-u(\lambda-i\epsilon) \Bigr]  \qquad (\lambda \in {\cal C}) \ .
\end{equation}
The loop equations are now obtained as follows. First we rewrite (\ref{scatt2}) as 
\begin{equation}
\frac{1}{2} W'(\lambda_a) = \frac{1}{N} \sum_{b\neq a} \frac{1}{\lambda_a - \lambda_b} \ , 
\end{equation}
multiply both sides by $\frac{1}{(\lambda_a-z)}$, and then sum over $a$, see for example  \cite{Marino:2004eq}. A short calculation\footnote{For the convenience of the reader, this is reproduced in Appendix~\ref{loop-eq}.} then leads to the so-called loop equation that the resolvent must obey, 
\be\label{loopeqs}
 u^2(z)-  W'(z) u(z) +  \frac{1}{N} u'(z) +  R(z) = 0 \ .
\ee
Here we have introduced the function $R(z)$, see eq.~(\ref{Req1}),  
\be\label{Req}
R(z) = \frac{1}{N} \sum_{a=1}^{N} \frac{ W'(\lambda_a) - W'(z)}{(\lambda_a - z)} = \sum_{i=1}^{n-1}\frac{\alpha_iu(z_i)}{z-z_i} \ . 
\ee
As the derivation in Appendix \ref{loop-eq} shows, the loop equations (\ref{loopeqs}) actually hold at any finite $N$. They are, moreover, in a form which is suitable for a large $N$ expansion. 
  
Next we introduce the so-called quantum corrected spectral curve, which is defined in terms of the resolvent via
\begin{equation}\label{qtm-spec}
y(z)=W'(z)-2u(z) \ .
\end{equation}
We can then rewrite (\ref{loopeqs}) as an equation that determines the spectral curve 
\be\label{yeq}
y^2(z) - \frac{2}{N} y'(z) = \bigl( W'(z) \bigr)^2 - \frac{2}{N} W''(z) - 4 R(z) \ . 
\ee
It is natural to view $y(z)dz$ as a differential on the underlying Riemann surface that emerges in the large $N$ limit, as we will see below. 

We should note that in terms of our original problem, the function $y(z)$ defined by (\ref{qtm-spec}) equals 
\begin{equation}\label{yGamma}
y(z)  =  \sum_{i=1}^{n-1} \frac{\alpha_i}{(z-z_i)}  - \frac{2}{N} \sum_{a=1}^{N} \frac{1}{(z-\lambda_a)}  =  \frac{1}{N} \frac{ \partial^2 \Gamma(z)}{\partial \Gamma(z)} = \frac{1}{N}  \partial \ln{\partial\Gamma}  , 
\end{equation}
where we have used the expression (\ref{partG}) for $\partial \Gamma(z)$, as well as the relation $N\alpha_i = (w_i-1)$ from eq.~(\ref{scatt2}). This identification (\ref{yGamma}), which will be important in what follows, is also true at finite $N$.
\medskip

For the rest of this subsection we shall concentrate on the leading large $N$ limit, and discuss what happens for finite $N$ in Section~\ref{sec:4.3}. Starting from eq.~(\ref{yeq}) let us replace $y(z)$ by its leading term in the large $N$ limit, which we denote by $y_0(z)$. Then (\ref{yeq}) reduces to 
\begin{equation}\label{specsol}
y_0^2(z)=  \bigl( W'(z) \bigr)^2 - 4 R_0(z) \ , 
\end{equation}
where we have dropped the terms with the explicit factors of $\frac{1}{N}$, and replaced the function $R(z)$ by its leading piece in the large $N$ limit, which we have denoted by $R_0(z)$; 
using (\ref{Req}) and (\ref{qtm-spec}) this can be written as 
\begin{equation}\label{R0}
R_0(z) =  \sum_{i=1}^{n-1}\frac{\alpha_iu_0(z_i)}{z-z_i} =   \frac{1}{2}\sum_{i=1}^{n-1}\frac{\alpha_i }{z-z_i} \, \Bigl(W'(z)-y_0(z)\Bigr)\Bigr|_{z\to z_i} \ .
\end{equation}
Together with the form of (\ref{potderiv}), this now allows us to rewrite eq.~(\ref{specsol}) as 
\begin{equation}\label{y0eq}
y_0^2(z) = \frac{{\tilde W}^2_{n-2}(z) - \prod_{i=1}^{n-1}(z-z_i)\tilde{R}_{n-3}(z)}{\prod_{i=1}^{n-1}(z-z_i)^2} 
\equiv  \frac{Q_{2n-4}(z)}{\prod_{i=1}^{n-1}(z-z_i)^2}    \ .
\end{equation}
Here $\tilde{W}_{n-2}(z)$ is a polynomial of degree $(n-2)$ that is determined by $W'(z)$, and  
$\tilde{R}_{n-3}(z)$ is similarly determined by $R_0(z)$ after rationalisation. Naively one may have thought that $\tilde{R}_{n-3}(z)$ is of degree $(n-2)$, but it actually has degree $(n-3)$  since the leading $\frac{1}{z}$ coefficient of $R_0(z)$ is proportional to $\sum_i\alpha_i u_0(z_i)$, which vanishes because of eq.~(\ref{wconstr}).

Note that while $\tilde{R}_{n-3}(z)$ implicitly depends on $y_0(z_i)$ via (\ref{R0}), we will think of its  $n-2$ parameters as initially unknown, i.e.\ not directly related to $y_0(z)$. Then we can think of eq.~(\ref{y0eq}) as determining $y_0(z)$ in terms of these unknown parameters. We will explain below (in Section~\ref{sec:indep}) how these parameters can subsequently be fixed by self-consistency. 

Returning to eq.~(\ref{y0eq}) we now note that the numerator defines a hyperelliptic curve with $(n-2)$ cuts (and genus $(n-3)$),
\be\label{hypellip}
\hat{y}^2(z)= Q_{2n-4}(z) = \bigl( {\tilde W}_{n-2}(z)\bigr)^2 - \prod_{i=1}^{n-1}(z-z_i)\tilde{R}_{n-3}(z) \equiv\alpha_n^2 \prod_{j=1}^{2(n-2)}(z-a_j) \ .
\ee
The one form $y_0(z)dz$ is then a meromorphic differential on this Riemann surface, and it has poles at $z=z_i$ with residue $\alpha_i$. We furthermore require that the residue at $z_n=\infty$ equals $\alpha_n$, and this then fixes the overall coefficient in (\ref{hypellip}).  
We also see from (\ref{qtm-spec}) that the leading order solution for the resolvent, $u_0(z) =\frac{1}{2}(W'(z)-y_0(z))$, has branch cuts and therefore discontinuities giving the eigenvalue densities as in (\ref{evdens}).  
An alternative form for $y_0(z)$ and the resolvent $u_0(z)$, exhibiting their single pole structure, is given in (\ref{spec-gen}) and (\ref{res-sol}), respectively.

\subsection{Independent parameters}\label{sec:indep}

In the usual matrix model treatment of this problem, the final step is to obtain  from the spectral curve the complete solution by 
determining the unknown polynomial $\tilde{R}_{n-3}(z)$ in (\ref{y0eq}), see the comment above eq.~(\ref{hypellip}). 
Often this is done by  
specifying the $(n-3)$ ``filling fractions" of the eigenvalue density along the $(n-3)$ independent A-cycles,\footnote{There are alternative conditions one could impose to determine these unknowns, such as demanding that eigenvalues do not tunnel, see \cite{Marino:2004eq}.} i.e.\ the $(n-3)$ periods 
\begin{equation}\label{fill-frac}
\frac{1}{4\pi i}\oint_{A_i}y_0(z)dz  = \int_{{\cal C}_i\equiv[a_{2i-1},a_{2i}]} d\lambda \,\rho(\lambda) = \nu_i \ ,  \qquad   i=1,\ldots , n-3 \ .
\end{equation}
Note that we actually only need to specify $(n-3)$ such periods since the overall coefficient of $\tilde{R}_{n-3}$ is fixed by the overall normalisation of $Q_{2n-4}(z)$ in (\ref{hypellip}). More specifically, the  coefficient of $z^{2n-4}$ in $({\tilde W}_{n-2}(z))^2$ equals $(\sum_{i=1}^{n-1}\alpha_i)^2= (2-\alpha_n)^2$, where we have used (\ref{sumalph}), and the $z^{n-3}$ coefficient of $\tilde{R}_{n-3}(z)$ is thus determined by the condition that the coefficient of $z^{2n-4}$ in $Q_{2n-4}(z)$ equals $\alpha_n^2$. 

As a consequence the $(n-3)$ conditions in (\ref{fill-frac}) are enough to solve for the polynomial 
$\tilde{R}_{n-3}(z)$, and this then also fixes the locations $a_j$ of the branch points of the cuts in (\ref{hypellip}). In this approach, the other set of periods (over the $B$-cycles) are determined in terms of the effective action of the matrix model 
\be\label{Bcycle}
\frac{1}{4\pi i}\oint_{B_i}y_0(z)dz  \sim \frac{\partial{\cal S}_{\rm eff}(\{ z_j, \nu_j\})}{\partial\nu_i} \ . 
\ee
Note that from the matrix model perspective, the $z_j$ are given {\it ab initio} --- they define the matrix model potential $W(z)$ as in (\ref{potential}). Thus we solve for the spectral curve and thus the eigenvalue density (including its support) in terms of the input data of the $(\alpha_i, z_i, \nu_i)$. 

As we will see more fully in Section~\ref{sec:spec-streb}, our present problem dictates a slightly different perspective: for us the $z_i$ are not specified initially, since they live on the auxiliary covering space which we are constructing by solving for the covering map. Instead, we will think of equations like (\ref{Bcycle}) as determining the $z_i$ in terms of the $B$-cycle periods.

More specifically, in our context it will be natural, instead, to  specify the $2(n-3)$ independent periods of the spectral curve --- over {\it both} the A- and B-cycles. We can then use these period integrals to determine the $(n-3)$ independent parameters of $\tilde{R}_{n-3}(z)$ together with the $(n-3)$ cross ratios of the $z_i$. (Recall that we have used the M\"obius invariance to fix three of the $z_i$). This is the sense in which our approach exhibits a slightly different angle and interpretation of the matrix model solution.

\subsection[The solution at finite \texorpdfstring{$N$}{N}]{\boldmath The solution at finite \texorpdfstring{$N$}{N}}\label{sec:4.3}

Having determined the spectral curve and thus the solution to the covering map to leading order in $N$, we will now study the solution at finite $N$; much of Sections~\ref{sec:spec-streb} and \ref{sec:6} will be independent of the considerations here, and readers can return to this subsection at a later stage if they wish. We start with the loop equation (\ref{yeq}) that was derived without taking the large $N$ limit (and therefore also holds at finite $N$)
\be\label{y1N}
y^2(z) - \frac{2}{N} y'(z) = \bigl( W'(z) \bigr)^2 - \frac{2}{N} W''(z) - 4 R(z) \ . 
\ee
It is striking that the combination that appears on the LHS is, in terms of the covering map, see eq.~(\ref{yGamma}), precisely
\begin{eqnarray}
y^2(z) - \frac{2}{N} y'(z)  & = &  \frac{1}{N^2} \Biggl[ \Bigl( \frac{\Gamma''}{\Gamma'} \Bigr)^2 - 2 \frac{\Gamma'''}{\Gamma'} + 2 \Bigl( \frac{\Gamma''}{\Gamma'} \Bigr)^2 \Biggr] \\
& = & - \frac{2}{ N^2} \Biggl[ \frac{\Gamma'''}{\Gamma'} - \frac{3}{2} \Bigl( \frac{\Gamma''}{\Gamma'} \Bigr)^2 \Biggr] = - \frac{2}{ N^2} S[\Gamma] \ ,  \label{schwloop}
\end{eqnarray}
where $S[\Gamma]$ is the Schwarzian derivative of $\Gamma(z)$. We furthermore rewrite the  RHS of (\ref{y1N}) in terms of the tree level curve $y_0(z)$ in (\ref{specsol}) as
\be\label{y0schw}
 - \frac{2}{ N^2} S[\Gamma]  = y_0^2(z) - \frac{2}{N} W''(z) - \frac{4}{N} R_1(z) \ ,
 \ee 
where $R(z) = R_0(z) +\frac{1}{N}R_1(z)$. Next we observe that the arguments leading up to (\ref{y0eq}) equally apply without taking the large $N$ limit --- in particular, the leading $\frac{1}{z}$ coefficient of $R(z)$ is proportional to $\sum_i\alpha_i u(z_i)$, and thus also vanishes because of eq.~(\ref{wconstr}) --- and we can therefore write (\ref{y1N}) as 
\be\label{ytild}
\tilde{y}^2(z)  \equiv y_0^2(z) - \frac{2}{N} W''(z) - \frac{4}{N} R_1(z) = \frac{\tilde{Q}_{2n-4}(z)}{\prod_{i=1}^{n-1}(z-z_i)^2} \ .
\ee
We can therefore interpret $\tilde{y}^2(z)$ is the $\frac{1}{N}$-corrected version of $y^2_0(z)$. Note that, because of the $- \frac{2}{N} W''(z)$ correction term, its ``residue" at $z_i$ is 
\be
\tilde{y}^2(z) \sim \frac{w_i^2-1}{N^2} \, \frac{1}{(z-z_i)^2} \ , 
\ee
whereas the coefficient of the double pole of $y^2(z)$ at $z=z_i$ equals $\frac{(w_i-1)^2}{N^2}$. This shift is required in order for it to agree with the Schwarzian of the covering map, see eq.~(\ref{y0schw}). The two correction terms in (\ref{ytild}) will also affect the periods (\ref{fill-frac}) and (\ref{Bcycle}). 
\smallskip

As we will see below in Section~\ref{sec:6},  $(-y_0^2)(z)dz^2$ defines a Strebel differential at leading order in $\frac{1}{N}$. As a consequence, the same is therefore true for the Schwarzian of the covering map. This is quite a remarkable relation between the Schwarzian derivative and Strebel differentials, and as far as we are aware, this connection had not been noticed before. Obviously, the Schwarzian derivative of the covering map can also be evaluated at finite $N$, and one could ask whether the Schwarzian at finite $N$ also defines a Strebel differential. Similarly, one could  analyse this question in the large $N$ limit, but including subleading $\frac{1}{N}$ corrections. It would be interesting to explore these directions further. 
\smallskip

Let us close this section by explaining how the matrix model approach leads to the correct number of solutions, see the discussion below (\ref{scatt}). As we have explained above in Section~\ref{sec:indep}, we can determine the leading order spectral curve $y_0(z)$, as well as the function $R_0(z)$, by specifying the periods. This argument also applies to the exact loop equation, and thus the periods determine the RHS of (\ref{y0schw}). This leads to a differential equation determining the covering map $\Gamma(z)$ which is of Sturm-Liouville type, see e.g.\ \cite{Schwrz}.  

At finite $N$, the LHS is the Schwarzian derivative of the covering map $\Gamma(z)$, and the equation therefore only determines $\Gamma(z)$ up to a M\"obius transformation in the $x$-space. This is a consequence of the transformation property of the Schwarzian \cite{Schwrz} 
\be\label{schwtransf}
S[f\circ g] = (S[f]\circ g) (g' )^2 + S[g] \ .
\ee
Indeed, if we take $g=\Gamma$ and let $f$ be a M\"obius transformation, we see that the Schwarzian is invariant under M\"obius transformations in the $x$-space,\footnote{This follows from $S[f]=0$. Note that this should not be confused with the transformation of the Schwarzian under a M\"obius transformation in the $z$-space, see eq.~(\ref{SGtrans}) below. } 
\be
S[f\circ \Gamma] = S[\Gamma] \ , \qquad \hbox{if $f$ is a M\"obius transformation.}
\ee
Thus (\ref{y0schw}) only allows us to determine the covering map up to M\"obius transformations in $x$, which means that we have the freedom to specify three of the branch points $x_i$ as expected. 

This should be contrasted with the leading order analysis, for which the loop equations determine  $y^2(z)$. This is related to the covering map as in (\ref{yGamma}), but the RHS of (\ref{yGamma}) is \emph{not} invariant under replacing $\Gamma \mapsto f\circ \Gamma$, where $f$ is a M\"obius transformation. Thus to leading order in $\frac{1}{N}$ the resulting solution would not allow us to specify three of the branch points $x_i$ at will. In retrospect, the fact that this subtlety is related to a $\frac{1}{N}$ effect is also clear from the original discussion below eq.~(\ref{scatt}): in that context it was important that there are actually only $(N-1)$, rather than $N$ independent scattering equations (which, in the large $N$ limit, is `a $\frac{1}{N}$ effect').

Finally we note that the RHS of (\ref{yGamma}) does not transform covariantly under M\"obius transformations, but it is only the combination $y^2(z)-\frac{2}{N}y'(z) = - \frac{2}{N^2}S[\Gamma](z)$ that appears in (\ref{y0schw}) which has a nice transformation behaviour. In fact, $\Gamma(z)$ transforms as a quadratic differential 
\be\label{SGtrans}
S[\Gamma(f(z))] = f'(z)^2 \, S[\Gamma(z)] \ , 
\ee
where $f(z)$ is a M\"obius transformation, and we have used (\ref{schwtransf}). Obviously, the difference between $y^2(z)$, and $y^2(z)-\frac{2}{N}y'(z)$, is subleading in $\frac{1}{N}$, and hence to leading order it does not matter which one considers. But the considerations of the last two paragraphs suggest that  it is more natural to include the $-\frac{2}{N}y'(z)$ correction and consider the LHS of  (\ref{y1N}) instead of just $y^2(z)$.

\section{The spectral curve and Feynman diagrams}\label{sec:spec-streb}

In the previous section we have outlined the method for solving the equations (\ref{scatt}) and hence for determining the branched covers for correlators with large twist. The result is encoded in the spectral curve in the form of eq.~(\ref{specsol}), which in turn determines the resolvent, see eq.~(\ref{qtm-spec}), and thereby also the discontinuities (``eigenvalue densities") see eq.~(\ref{evdens}).  The branched covers themselves are then obtained via eq.~(\ref{yGamma}), or more accurately via eq.~(\ref{y0schw}), see the discussion at the end of  Section~\ref{sec:4.3}. 

As we discussed at the end of Section~\ref{sec:indep} it is natural to fix the independent parameters (including the $z_i$) by specifying the period integrals over both the $A$- and the $B$-cycles. It is the aim of this section to explain in more detail why this is so. In the process of doing so we will also exhibit the meaning of the periods themselves. 

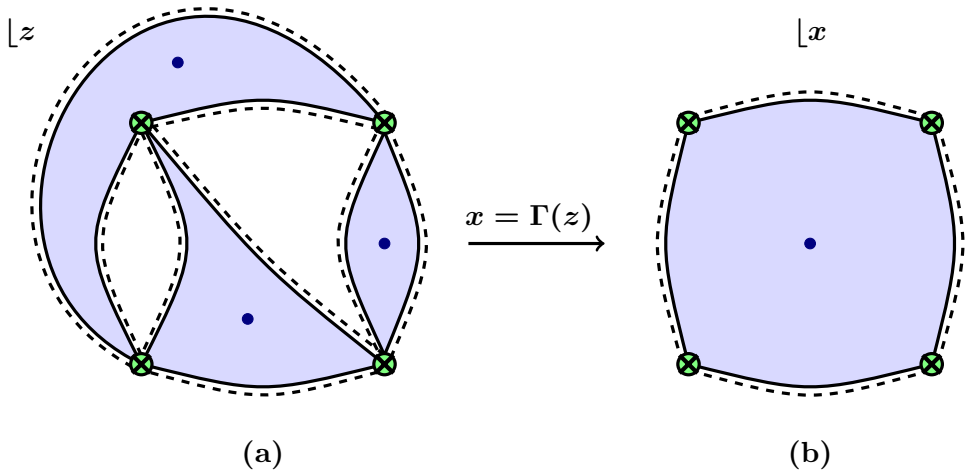
\begin{figure}[htb]
\vspace*{-2.5cm}
\hspace*{-2.1cm}
\centering
\begin{tikzpicture}[scale = 0.8]

%WORLDSHEET (z co-ordinates)

%Solid Lines

\draw[black, very thick, fill=blue!15!] (4,4)..controls(0,9) and (-4,2)..(0,0)..controls(-1,2)..(0,4)..controls (2,4.5)..(4,4);

\draw[black, very thick, fill=blue!15!] (0,0)..controls (2,-0.5)..(4,0)..controls (2,1.6)..(0,4)..controls (1,2)..(0,0);

\draw[black, very thick, fill=blue!15!] (4,0.15)..controls (3.15,2)..(4,3.85)..controls (4.75,2)..(4,0.15);

%Dashed Lines

\draw[black, very thick, dashed] (0,0.15)..controls (-0.85,2)..(0,3.85)..controls (0.85,2)..(0,0.15);

\draw[black, very thick, dashed] (4,0)..controls (3,2)..(3.9,3.9);

\draw[black, very thick, dashed] (4,0.15)..controls (2,1.75)..(0.15,4);

\draw[black, very thick, dashed] (0.15,3.93)..controls (2,4.35)..(3.9,3.9);

\draw[black, very thick, dashed] (-0.1,-0.07)..controls (2,-0.65)..(4,-0.1);

\draw[black, very thick, dashed] (4.1,0.1)..controls (4.9,2)..(4,4.1);

\draw[black, very thick, dashed] (4.07,4.1)..controls (0,9.15) and (-4.25,2.15)..(-0.1,-0.07);

% Four twist insertions
\filldraw [color=black, fill=green!50, very thick] (4,4) circle (5pt) node[cross,black] {};
\filldraw [color=black, fill=green!50, very thick] (0,0) circle (5pt) node[cross,black] {};
\filldraw [color=black, fill=green!50, very thick]  (0,4) circle (5pt) node[cross,black] {};
\filldraw [color=black, fill=green!50, very thick]  (4,0) circle (5pt) node[cross,black] {};

%Poles of the Covering Map
\filldraw[blue!50!black!100!] (4,2) circle (2.5pt);
\filldraw[blue!50!black!100!] (1.75,0.75) circle (2.5pt);
\filldraw[blue!50!black!100!] (0.6,5) circle (2.5pt);

% The Covering Map
\draw[black, very thick,->] (5.38,2)--(7.62,2) node[anchor=south east] {$\boldsymbol{x=\Gamma(z)}$};

% SPACETIME (x co-ordinates)

\draw[black, very thick, fill=blue!15!] (9,0)..controls (11,-0.5)..(13,0)..controls (13.5,2)..(13,4)..controls (11,4.5)..(9,4)..controls (8.5,2)..(9,0);

\draw[black, very thick, dashed] (9,-0.1)..controls (11,-0.65)..(13,-0.1);

\draw[black, very thick, dashed] (13.1,0)..controls (13.65,2)..(13.1,4);

\draw[black, very thick, dashed] (13,4.1)..controls (11,4.65)..(9,4.1);

\draw[black, very thick, dashed] (8.9,4)..controls (8.35,2)..(8.9,0);

% Four twist insertions
\filldraw [color=black, fill=green!50, very thick] (13,4) circle (5pt) node[cross,black] {};
\filldraw [color=black, fill=green!50, very thick] (9,0) circle (5pt) node[cross,black] {};
\filldraw [color=black, fill=green!50, very thick] (13,0) circle (5pt) node[cross,black] {};
\filldraw [color=black, fill=green!50, very thick] (9,4) circle (5pt) node[cross,black] {};

%Infinity
\filldraw[blue!50!black!100!] (11,2) circle (2.5pt);

%Specifying Co-ordinates
\node[] at (-2, 5.5)  {$\boldsymbol{\lfloor z}$};
\node[] at (11, 5.5)  {$\boldsymbol{\lfloor x}$};

%Specifying (a) and (b)
\node[] at (2, -1.5) {\textbf{(a)}};
\node[] at (11, -1.5) {\textbf{(b)}};
\end{tikzpicture}
\caption{An illustration of the Feynman graph of a symmetric orbifold correlator, constructed via the preimage of a Jordan curve under $\Gamma$. The case with $n=4$ and all $w_i=2$ is depicted, for which $N=3$.}
\label{fig:2}
\end{figure}

Let us begin by describing the covering map using the diagrammatic picture of \cite{Pakman:2009zz}. They associate to each covering map a diagram by considering a Jordan curve  passing through the $n$ points $x_i$ (in some prescribed fixed ordering) on the spacetime sphere, and enclosing $x=\infty$.\footnote{More precisely, they actually introduce a  ``bifundamental" double line curve, with the inner solid curve enclosing $x=\infty$, and an outer dashed curve, see Fig.~\ref{fig:2}.} This curve then has a pre-image in the covering space, where it defines a graph with the vertices being the branch points $z_i$ with $\Gamma(z_i)=x_i$. The poles $\lambda_a$ of the covering map $\Gamma(z)$  are, on the other hand, the pre-images of $x=\infty$, and since the Jordan curve encloses $x=\infty$, they are associated with the $N$ faces (or so-called ``coloured loops") of the resulting configuration. 

The complement of the above Jordan curve defines another set of $N$ faces (the so-called ``dashed loops''). The construction of \cite{Pakman:2009zz} therefore associates to each covering map a Feynman-like double line diagram comprising of ``Wick contractions" between the $n$ vertices, see Fig.~\ref{fig:2}. These diagrams are the analogue of the free field 't~Hooft diagrams for $n$-point correlators in Yang-Mills theory.\footnote{Note that in the present case we have $2w_i$ double lines emerging from each vertex with twist $w_i$.} As in that case, we associate a genus to the diagram which is that of the covering space that is triangulated by the above faces.  

Since we are considering here genus zero covering spaces, our double-line diagrams are planar. 
Following \cite{Gopakumar:2004qb} we associate to each such allowed diagram a {\it skeleton graph} whereby we glue all homotopically equivalent edges. It is easy to verify, using Euler's formula for genus zero, 
\be
n_V - n_E + n_F = 2 \ , 
\ee
that the resulting skeleton graph ${\cal G}$ will have $n_E=(3n-6)$ edges and $n_F=(2n-4)$ faces (which are generically triangular), as well as obviously $n_V=n$ vertices. It will also be useful to consider the dual graph ${\cal G}^D$ which then has $n$ faces, $(2n-4)$ vertices and $(3n-6)$ dual edges which are transverse to the edges of the original skeleton graph. 

Given that we have `collapsed' homotopic Wick contractions, a given skeleton graph ${\cal G}$ accounts, however, for many inequivalent double-line diagrams. The \emph{additional} data that is needed to reconstruct the double-line diagram from the skeleton graph is simply the number of edges between pairs of vertices $(i,j)$ in the original double-line diagram, which we denote by $n_{ij}=n_{ji}$. These integers $n_{ij}$ are constrained only by positivity and the requirement that at each vertex (labelled by $i$) they satisfy
\begin{equation}\label{nconstr}
\sum_{j\neq i}n_{ij}= 2 w_i \ ,  \qquad (\forall i=1, \ldots, n) \ .
\end{equation}
Since there can only be $(3n-6)$ edges (and therefore as many non-zero $n_{ij}$), and taking into account the $n$ constraints from (\ref{nconstr}), we see that only $(2n-6)$ of the $n_{ij}$ are independent. 
So far, everything we have said is true for finite $w_i$. 
\medskip

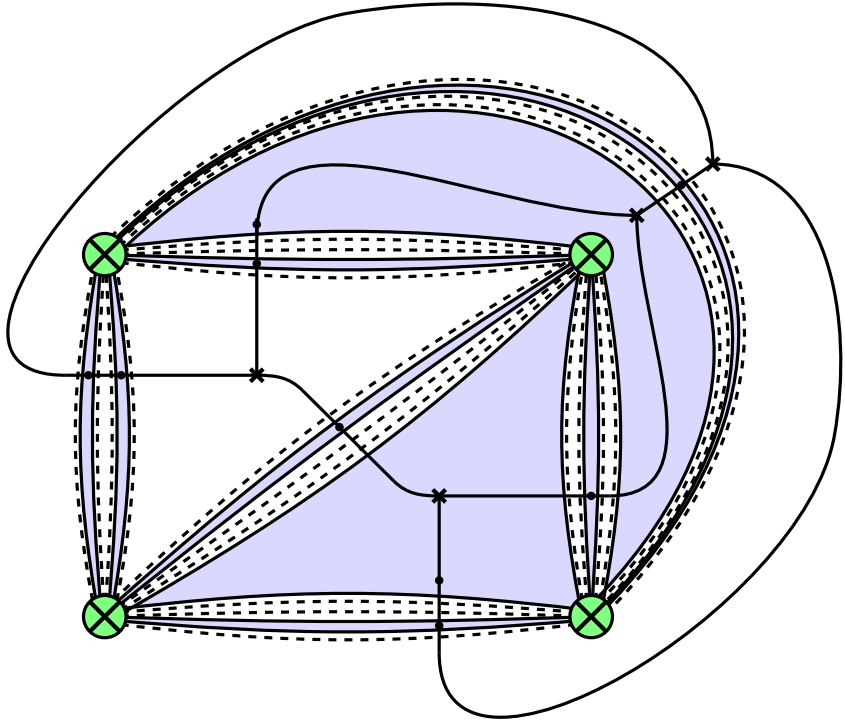
\begin{figure}[htb]
\vspace*{-2cm}
% \hspace*{-1.5cm}
\centering
\begin{tikzpicture}[scale = 0.8]

%filling in the light blue areas

\fill[color = blue!15!] (-4,-3 + 0.10) to[out = 7, in = 180 - 7] (4,-3 + 0.10) --  (4 - 0.15,-3) to[out = 90 + 11, in = -90 - 11] (4 - 0.15,3) -- (4 + 0.10/1.44,3 - 0.10/1.44) to[out = 180 + 37 + 7, in = 37 - 7] (-4 + 0.10/1.44,-3 - 0.10/1.44);
\fill[color = blue!15!] (-4,3 + 0.10) to[out = 7, in = 180 - 7] (4,3 + 0.10) -- (4 + 0.15,3) to[in = 90 - 11, out = -90 + 11] (4 + 0.15,-3) -- (4 - 0.1,-3 + 0.1) to[in = 45, out = 45, distance = 8.4cm] (-4 + 0.1,3 - 0.1);
\fill[color = blue!15!] (-4,-3) to[out = -2, in = -180 + 2] (4,-3) -- (4,-3 - 0.05) to[in = -5, out = -180 + 5] (-4,-3 - 0.05);
\fill[color = blue!15!] (-4,3) to[out = -2, in = -180 + 2] (4,3) -- (4,3 - 0.05) to[in = -5, out = -180 + 5] (-4,3 - 0.05);
\fill[color = blue!15!] (4,-3) to[out = 90 - 4, in = -90 + 4] (4,3) -- (4,3) to[in = 90 + 4, out = -90 - 4] (4,-3);
\fill[color = blue!15!] (-4 + 0.05,-3) to[out = 90 - 5, in = -90 + 5] (-4 + 0.05,3) -- (-4 + 0.10,3) to[in = 90 - 10, out = -90 + 10] (-4 + 0.10,-3);
\fill[color = blue!15!] (-4 - 0.05,-3) to[out = 90 + 5, in = -90 - 5] (-4 - 0.05,3) -- (-4 - 0.10,3) to[in = 90 + 10, out = -90 - 10] (-4 - 0.10,-3);
\fill[color = blue!15!] (-4,-3) to[out = 37 + 2, in = 180 + 37 - 2] (4,3) -- (4 - 0.05/1.44,3 + 0.05/1.44) to[in = 37 + 5, out = 180 + 37 - 5] (-4 - 0.05/1.44,-3 + 0.05/1.44);
\fill[color = blue!15!] (-4 - 0.02,3 + 0.02) to[out = 45, in = 45, distance = 8.95cm] (4 + 0.02,-3 - 0.02) -- (4 + 0.05,-3 - 0.05) to[in = 45, out = 45, distance = 9.15cm] (-4 - 0.05,3 + 0.05);

%dual strebel graph

\draw[very thick] (1.5,-3.6) -- (1.5,-1);
\draw[very thick] (-1.5,1) -- (-1.5,3.6);
\draw[very thick] (1.5,-1) -- (4.3,-1);
\draw[very thick] (-4.65,1) -- (-1.5,1);
\draw[very thick] (-0.75,0.75) -- (0.75,-0.75);
\draw[very thick] (-0.75,0.75) to[out = 135, in = 0] (-1.5,1);
\draw (-1.5,1) node[cross=4] {};
\draw[very thick] (0.75,-0.75) to[out = -45, in = 180] (1.5,-1);
\draw (1.5,-1) node[cross=4] {};
\draw[very thick] (4.3,-1) to[out = 0, in = -90] (4.75,3.65);
\draw[very thick] (-1.5,3.3) to[out = 90, in = 180] (4.75,3.65);
\draw (4.75,3.65) node[cross=4] {};
\draw[very thick] (4.75,3.65) -- (6,4.5);
\draw[very thick] (-4.65,1) to[out = 180, in = 190] (0,7) to[out = 10, in = 90] (6,4.5);
\draw[very thick] (1.5,-3.6) to[out = -90, in = -100] (8,0) to[out = 80, in = 0] (6,4.5);
\draw (6,4.5) node[cross=4] {};

% bottom fatgraph propagators and poles

\draw[very thick, black] (-4,-3) to[out = -2, in = -180 + 2] (4,-3);
\draw[very thick, black, dashed] (-4,-3) to[out = 2, in = 180 - 2] (4,-3);
\draw[very thick, black] (-4,-3 - 0.05) to[out = -5, in = -180 + 5] (4,-3 - 0.05);
\draw[very thick, black, dashed] (-4,-3 + 0.05) to[out = 5, in = 180 - 5] (4,-3 + 0.05);
\draw[very thick, black, dashed] (-4,-3 - 0.10) to[out = -7, in = -180 + 7] (4,-3 - 0.10);
\draw[very thick, black] (-4,-3 + 0.10) to[out = 7, in = 180 - 7] (4,-3 + 0.10);
\fill (1.5,-2.4) circle (0.07);
\fill (1.5,-3.15) circle (0.07);

%top fatgraph propagators and poles

\draw[very thick, black] (-4,3) to[out = -2, in = -180 + 2] (4,3);
\draw[very thick, black, dashed] (-4,3) to[out = 2, in = 180 - 2] (4,3);
\draw[very thick, black] (-4,3 - 0.05) to[out = -5, in = -180 + 5] (4,3 - 0.05);
\draw[very thick, black, dashed] (-4,3 + 0.05) to[out = 5, in = 180 - 5] (4,3 + 0.05);
\draw[very thick, black, dashed] (-4,3 - 0.10) to[out = -7, in = -180 + 7] (4,3 - 0.10);
\draw[very thick, black] (-4,3 + 0.10) to[out = 7, in = 180 - 7] (4,3 + 0.10);
\fill (-1.5,2.85) circle (0.07);
\fill (-1.5,3.5) circle (0.07);

% right fatgraph propagators and poles

\draw[very thick, black] (4,-3) to[out = 90 - 4, in = -90 + 4] (4,3);
\draw[very thick, black] (4,-3) to[out = 90 + 4, in = -90 - 4] (4,3);
\draw[very thick, black, dashed] (4 + 0.05,-3) to[out = 90 - 5, in = -90 + 5] (4 + 0.05,3);
\draw[very thick, black, dashed] (4 - 0.05,-3) to[out = 90 + 5, in = -90 - 5] (4 - 0.05,3);
\draw[very thick, black, dashed] (4 + 0.10,-3) to[out = 90 - 10, in = -90 + 10] (4 + 0.10,3);
\draw[very thick, black, dashed] (4 - 0.10,-3) to[out = 90 + 10, in = -90 - 10] (4 - 0.10,3);
\draw[very thick, black] (4 + 0.15,-3) to[out = 90 - 11, in = -90 + 11] (4 + 0.15,3);
\draw[very thick, black] (4 - 0.15,-3) to[out = 90 + 11, in = -90 - 11] (4 - 0.15,3);
\fill (4,-1) circle (0.07);

%left fatgraph propagators and poles

\draw[very thick, black, dashed] (-4,-3) to[out = 90 - 4, in = -90 + 4] (-4,3);
\draw[very thick, black, dashed] (-4,-3) to[out = 90 + 4, in = -90 - 4] (-4,3);
\draw[very thick, black] (-4 + 0.05,-3) to[out = 90 - 5, in = -90 + 5] (-4 + 0.05,3);
\draw[very thick, black] (-4 - 0.05,-3) to[out = 90 + 5, in = -90 - 5] (-4 - 0.05,3);
\draw[very thick, black] (-4 + 0.10,-3) to[out = 90 - 10, in = -90 + 10] (-4 + 0.10,3);
\draw[very thick, black] (-4 - 0.10,-3) to[out = 90 + 10, in = -90 - 10] (-4 - 0.10,3);
\draw[very thick, black, dashed] (-4 + 0.15,-3) to[out = 90 - 11, in = -90 + 11] (-4 + 0.15,3);
\draw[very thick, black, dashed] (-4 - 0.15,-3) to[out = 90 + 11, in = -90 - 11] (-4 - 0.15,3);
\fill (-4.27,1) circle (0.07);
\fill (-3.73,1) circle (0.07);

%diagonal fatgraph propagators and poles

\draw[very thick, black] (-4,-3) to[out = 37 + 2, in = 180 + 37 - 2] (4,3);
\draw[very thick, black, dashed] (-4,-3) to[out = 37 - 2, in = 180 + 37 + 2] (4,3);
\draw[very thick, black] (-4 - 0.05/1.44,-3 + 0.05/1.44) to[out = 37 + 5, in = 180 + 37 - 5] (4 - 0.05/1.44,3 + 0.05/1.44);
\draw[very thick, black, dashed] (-4 + 0.05/1.44,-3 - 0.05/1.44) to[out = 37 - 5, in = 180 + 37 + 5] (4 + 0.05/1.44,3 - 0.05/1.44);
\draw[very thick, black, dashed] (-4 - 0.10/1.44,-3 + 0.10/1.44) to[out = 37 + 7, in = 180 + 37 - 7] (4 - 0.10/1.44,3 + 0.10/1.44);
\draw[very thick, black] (-4 + 0.10/1.44,-3 - 0.10/1.44) to[out = 37 - 7, in = 180 + 37 + 7] (4 + 0.10/1.44,3 - 0.10/1.44);
\fill (-0.14,0.14) circle (0.07);

%the difficult fatgraph propagators and poles

\draw[very thick, black] (-4 - 0.02,3 + 0.02) to[out = 45, in = 45, distance = 8.95cm] (4 + 0.02,-3 - 0.02);
\draw[very thick, black, dashed] (-4,3) to[out = 45, in = 45, distance = 8.8cm] (4,-3);
\draw[very thick, black] (-4 - 0.05,3 + 0.05) to[out = 45, in = 45, distance = 9.15cm] (4 + 0.05,-3 - 0.05);
\draw[very thick, black, dashed] (-4 + 0.05,3 - 0.05) to[out = 45, in = 45, distance = 8.55cm] (4 - 0.05,-3 + 0.05);
\draw[very thick, black, dashed] (-4 - 0.1,3 + 0.1) to[out = 45, in = 45, distance = 9.3cm] (4 + 0.1,-3 - 0.1);
\draw[very thick, black] (-4 + 0.1,3 - 0.1) to[out = 45, in = 45, distance = 8.4cm] (4 - 0.1,-3 + 0.1);
\fill (5.48,4.15) circle (0.07);

%strebel zeroes

\foreach \x in {0,1}
\foreach \y in {0,1}
{
\draw[very thick, fill = green!50] (-4 + 8*\x, -3 + 6*\y) circle (0.35);
\draw[ultra thick] (-4 + 8*\x + 0.35*0.18/0.25, -3 + 6*\y + 0.35*0.18/0.25) -- (-4 + 8*\x - 0.35*0.18/0.25, -3 + 6*\y - 0.35*0.18/0.25);
\draw[ultra thick] (-4 + 8*\x + 0.35*0.18/0.25, -3 + 6*\y - 0.35*0.18/0.25) -- (-4 + 8*\x - 0.35*0.18/0.25, -3 + 6*\y + 0.35*0.18/0.25);
% \fill (-4 + 8*\x, -3 + 6*\y) circle (0.05);
}
\end{tikzpicture}
\vspace*{-1cm}
\caption{The Feynman graph for a four-point correlator with $w_i=5$ and therefore $N = 9$. The critical points are denoted by $\otimes$. The dual of its skeleton graph, ${\cal G}^D$, is described by the black solid lines (and its vertices are denoted by crosses); it corresponds to the graph of critical horizontal trajectories of the Strebel differential as discussed in Section~\ref{sec:6}.}
\label{fig:3}
\end{figure}

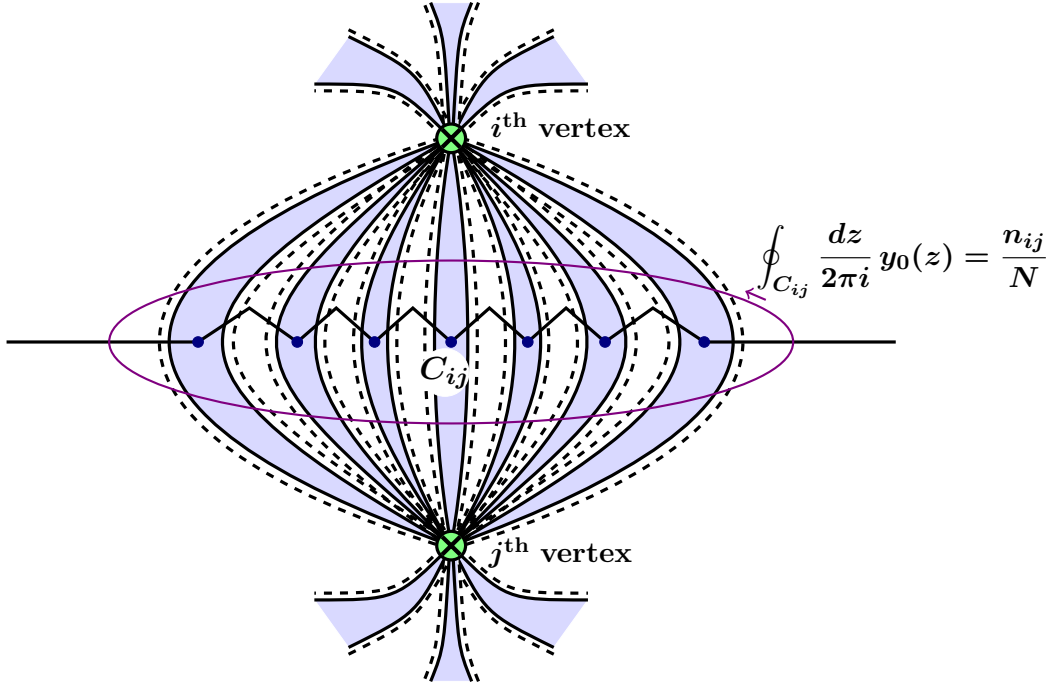
\begin{figure}[htb]
\hspace*{1cm}
\centering
\begin{tikzpicture}[scale = 0.9]
%Middle rivers

%3rd
\draw[very thick, fill=blue!15!] (0,0) to[out = 20, in = -20,  looseness=2.5] (0,6) to[out=-30, in=30, looseness=2.2] (0,0);

\draw[very thick, dashed] (0,0-0.05) to[out = 20-3, in = -20+3,  looseness=2.5] (0,6+0.05) (0,6-0.05) to[out=-30-3, in=30+3, looseness=2.2] (0,0+0.05);

\draw[very thick, fill=blue!15!] (0,0) to[out = 160, in = -160,  looseness=2.5] (0,6) to[out=-150, in=150,looseness=2.2] (0,0);

\draw[very thick, dashed] (0,0-0.05) to[out = 160+3, in = -160-3,  looseness=2.5] (0,6+0.05) (0,6-0.05) to[out=-150+3, in=150-3, looseness=2.2] (0,0+0.05);

%2nd

\draw[very thick, fill=blue!15!] (0,0) to[out = 40, in = -40,  looseness=1.9] (0,6) to[out=-45, in=45, looseness=1.6] (0,0);

\draw[very thick, dashed] (0,0-0.05) to[out = 40-3, in = -40+3,  looseness=1.9] (0,6+0.05) (0,6-0.05) to[out=-45-3, in=45+3, looseness=1.6] (0,0+0.05);

\draw[very thick, fill=blue!15!] (0,0) to[out = 180-40, in = -180+40,  looseness=1.9] (0,6) to[out=-180+45, in=180-45, looseness=1.6] (0,0);

\draw[very thick, dashed] (0,0-0.05) to[out = 180-40+3, in = -180+40-3,  looseness=1.9] (0,6+0.05) (0,6-0.05) to[out=-180+45+3, in=180-45-3, looseness=1.6] (0,0+0.05);

%1st
\draw[very thick, fill=blue!15!] (0,0) to[out = 55, in = -55,  looseness=1.3] (0,6) to[out=-58, in=58, looseness=1] (0,0);

\draw[very thick, dashed] (0,0-0.05) to[out = 55-3, in = -55+3,  looseness=1.3] (0,6+0.05) (0,6-0.05) to[out=-58-3, in=58+3, looseness=0.95] (0,0+0.05);

\draw[very thick, fill=blue!15!] (0,0) to[out = 180-55, in = -180+55,  looseness=1.3] (0,6) to[out=-180+58, in=180-58, looseness=1] (0,0);

\draw[very thick, dashed] (0,0-0.05) to[out = 180-55+3, in = -180+55-3,  looseness=1.3] (0,6+0.05) (0,6-0.05) to[out=-180+58+3, in=180-58-3, looseness=0.95] (0,0+0.05);

%Central color loop (0th river)
\draw[very thick, fill=blue!15!] (0,0) to[out = 80, in = -80,  looseness=0.8] (0,6) to[out=-180+80, in=180-80, looseness=0.8] (0,0);

\draw[very thick, dashed] (0+0.05,0+0.05) to[out = 80-3, in = -80+3,  looseness=0.85] (0+0.05,6) (0-0.05,6) to[out=-180+80-3, in=180-80+3, looseness=0.85] (0-0.05,0);

%Upper rivers

%left
\draw[color=white, fill=blue!15!] (0,6)..controls (-0.8,6.8)..(-2,6.8)--(-1.5,7.5)..controls (-0.5,7)..(0,6);
\draw[very thick] (0,6)..controls (-0.8,6.8)..(-2,6.8) (-1.5,7.5)..controls (-0.5,7)..(0,6);
\draw[very thick, dashed] (-0.1,6)..controls (-0.8,6.7)..(-2,6.7) (-1.5,7.6)..controls (-0.5,7.1)..(0.1,6);

%right
\draw[color=white, fill=blue!15!] (0,6)..controls (0.8,6.8)..(2,6.8)--(1.5,7.5)..controls (0.5,7)..(0,6);
\draw[very thick] (0,6)..controls (0.8,6.8)..(2,6.8) (1.5,7.5)..controls (0.5,7)..(0,6);
\draw[very thick, dashed] (0.1,6)..controls (0.8,6.7)..(2,6.7) (1.5,7.6)..controls (0.5,7.1)..(-0.1,6);

%central
\draw[color=white, fill=blue!15!] (-0.02,6)..controls (-0.1,7.3)..(-0.3,8)--(0.3,8)..controls (0.1,7.3)..(0.1,6)--(0.02,6);
\draw[very thick] (-0.02,6)..controls (-0.1,7.5)..(-0.3,8) (0.3,8)..controls (0.1,7.5)..(0.02,6)--(0,6);
\draw[very thick, dashed] (-0.1,6)..controls (-0.2,7.5)..(-0.4,8) (0.4,8)..controls (0.2,7.5)..(0.1,6);

%Lower Rivers

%left
\draw[color=white, fill=blue!15!] (0,6-6)..controls (-0.8,-6.8+6)..(-2,-6.8+6)--(-1.5,-7.5+6)..controls (-0.5,-7+6)..(0,-6+6);
\draw[very thick] (0,6-6)..controls (-0.8,-6.8+6)..(-2,-6.8+6) (-1.5,-7.5+6)..controls (-0.5,-7+6)..(0,-6+6);
\draw[very thick, dashed] (-0.1,-6+6)..controls (-0.8,-6.7+6)..(-2,-6.7+6) (-1.5,-7.6+6)..controls (-0.5,-7.1+6)..(0.1,-6+6);

%right
\draw[color=white, fill=blue!15!] (0,-6+6)..controls (0.8,-6.8+6)..(2,-6.8+6)--(1.5,-7.5+6)..controls (0.5,-7+6)..(0,-6+6);
\draw[very thick] (0,-6+6)..controls (0.8,-6.8+6)..(2,-6.8+6) (1.5,-7.5+6)..controls (0.5,-7+6)..(0,-6+6);
\draw[very thick, dashed] (0.1,-6+6)..controls (0.8,-6.7+6)..(2,-6.7+6) (1.5,-7.6+6)..controls (0.5,-7.1+6)..(-0.1,-6+6);

%central
\draw[color=white, fill=blue!15!] (-0.02,-6+6)..controls (-0.1,-7.3+6)..(-0.3,-8+6)--(0.3,-8+6)..controls (0.1,-7.3+6)..(0.1,-6+6)--(0.02,-6+6);
\draw[very thick] (-0.02,-6+6)..controls (-0.1,-7.5+6)..(-0.3,-8+6) (0.3,-8+6)..controls (0.1,-7.5+6)..(0.02,-6+6)--(0,-6+6);
\draw[very thick, dashed] (-0.1,-6+6)..controls (-0.2,-7.5+6)..(-0.4,-8+6) (0.4,-8+6)..controls (0.2,-7.5+6)..(0.1,-6+6);

%Edge of the dual graph
\draw[very thick] (-6.5,3)--(-3.7,3)--(-2.95,3.5)--(-2.25,3)--(-1.68,3.5)--(-1.12,3)--(-0.56,3.5)--(0,3)--(0.56,3.5)--(1.12,3)--(1.68,3.5)--(2.25,3)--(2.95,3.5)--(3.7,3)--(6.5,3);

%Poles of the Covering Map
\filldraw[blue!50!black!100!] (0,3) circle (2.2pt) (1.12,3) circle (2.2pt)   (2.25,3) circle (2.2pt) (3.7,3) circle (2.2pt) (-1.12,3) circle (2.2pt) (-2.25,3) circle (2.2pt) (-3.7,3) circle (2.2pt);

%Contour
\draw[thick, violet] (0,3) ellipse (5cm and 1.2cm);

% Two twist insertions
\filldraw [color=black, fill=green!50, very thick] (0,0) circle (6pt) node[cross,black] {};
\filldraw [color=black, fill=green!50, very thick] (0,6) circle (6pt) node[cross,black] {};

%Nodes
\node at (1.6, -0.1)  {$\boldsymbol{j^{\text{\textbf{th}}}}$ \textbf{vertex}};
\node at (1.6, 6.2)  {$\boldsymbol{i^{\text{\textbf{th}}}}$ \textbf{vertex}};
\node at (6.6, +4.2)  {\scalebox{1}{$\boldsymbol{\displaystyle\oint_{C_{ij}} \frac{dz}{2\pi i}\, y_0(z)=\frac{n_{ij}}{N}}$}};

\draw[color=white, fill=white] (-0.09, 2.55) circle (0.35);
\node at (-0.09, 2.55)  {\scalebox{1.1}{$\boldsymbol{C_{ij}}$}};

%One arrow
\draw[->, thick, violet] (4.6,3.62)--(4.3,3.75);

\end{tikzpicture}
\caption{The period integral along the cut that corresponds to the dual edge $\widehat{(ij)}$.}
\label{fig:final}
\end{figure}

We are interested in the regime where we scale the twists as $w_i \sim \alpha_i N$ with $N$ large, and then the $n_{ij}$, when they are non-zero, also generically scale as $N$. We note in passing that the number of covering maps scales in this limit as 
\be\label{numbergen}
\# (\hbox{branched coverings}) \sim N^{2n-6} \ , 
\ee
as follows from the argument below eq.~(\ref{nconstr}); for $n=4$, this is in agreement with eq.~(\ref{number}). 
In the original double-line diagram we had one pole $\lambda_a$ for each of the $N$ coloured faces, but in the skeleton graph only $(2n-4)$ (generically triangular) faces remain. This implies that, at large $N$, most of the  $N$ poles are associated with the two-edged faces formed from homotopic Wick contraction, i.e.\ with the faces that disappear when we glue the double-line diagram to form the skeleton graph, see Fig.~\ref{fig:3}. In fact, as we have learnt in Section~\ref{sec:MM}, these poles coalesce in the large $N$ limit into a system of cuts ${\cal C}$, which are transverse to the original edges, and are now seen to build up the edges of the dual skeleton graph ${\cal G}^D$. Thus we can identify the different cuts in the cut-system ${\cal C}$ with the edges of the dual skeleton graph ${\cal G}^D$, and the $(2n-4)$ vertices of ${\cal G}^D$ with the end-points of the cuts, i.e.\ the $(2n-4)$ $a_j$ from eq.~(\ref{hypellip}). Furthermore, the number of poles associated to the dual edge $\widehat{(ij)}$, i.e.\ the edge of ${\cal G}^D$ that is transverse to the edge $(ij)$ of ${\cal G}$, is approximately $\frac{n_{ij}}{2}$ in the leading large $N$ limit.\footnote{Note that, at large $N$, only approximately half of the Wick contractions $n_{ij}$ correspond to poles in the coloured loops, see Fig.~\ref{fig:final}.}   Finally, the $n$ faces of ${\cal G}^D$ each contain one of the $n$ vertices $z_i$ of ${\cal G}$; these can be identified with the simple poles of the spectral curve $y_0(z)$ as in (\ref{spec-fin}).

It should now be clear how the matrix model solution characterised by the spectral curve $y_0(z)$, is related to the covering map described in terms of the skeleton graph ${\cal G}$ and the numbers $n_{ij}$. Starting from the spectral curve, we identify the cut system ${\cal C}$ with the dual skeleton graph ${\cal G}^D$. The discontinuity of $y_0(z)$ across a cut counts the fraction of $\lambda_a$ poles associated to this cut,  and thus the period integral of $y_0(z)$ along this cut is exactly twice the fraction $\frac{n_{ij}}{2N}$ one associates to the corresponding dual edge $\widehat{(ij)}$ of ${\cal G}^D$, see eqs.~(\ref{qtm-spec}), (\ref{evdens}) and Fig.~\ref{fig:final}. As we have seen above, see eq.~(\ref{nconstr}),  there are precisely $(2n-6)$ independent such $n_{ij}$ from the viewpoint of ${\cal G}^D$. In terms of the spectral curve, this follows from the fact that the integral of the spectral curve $y_0(z)$ around the edges of a given face of ${\cal G}^D$, say the one that contains $z_i$, equals the residue of $y_0(z)$ at $z_i$, i.e.\ $\alpha_i = \frac{w_i}{N}$. This leads to $n$ constraints among the  $(3n-6)$ integrals of $y_0(z)$ along the edges of ${\cal G}^D$ (or the cuts of the cut-system ${\cal C}$), and hence reduces the number of independent period integrals to $2n-6$. If we denote an appropriate set of $(2n-6)$ independent $n_{ij}$ by $n^{(l)}$ and $\tilde{n}^{(l)}$, with $l=1,\ldots, n-3$, we have\footnote{Since $y_0(z)$ is meromorphic with simple poles at $z_i$ (\ref{resolvent}), we have to consider the nontrivial homology cycles of the hyperelliptic curve with $n$ punctures. Since the residues around the poles are fixed ($=\alpha_i$), we have a choice in picking $(2n-6)$ independent periods. This precisely corresponds to the freedom of picking different independent $n_{ij}$.} 
\begin{equation}\label{periods}
\frac{1}{4\pi i}\oint_{A_l}y_0(z) dz \equiv \nu_l =\frac{n^{(l)} }{2N}\ ,  \qquad \frac{1}{4\pi i}\oint_{B_l} y_0(z) dz \equiv \mu_l =\frac{\tilde{n}^{(l)}}{2 N}  \ .
\end{equation}
This therefore determines the dual skeleton graph ${\cal G}^D$ and the ratios $\frac{n_{ij}}{N}$ in terms of the spectral curve.

Conversely, the ratios $\frac{n_{ij}}{N}$ along the edges of the dual skeleton graph ${\cal G}^D$  fix the periods of the spectral curve, and hence determine it by the arguments of the previous section. Note that the discrete family of covering maps,  labelled by the different double-line diagrams, goes over, in the large twist limit, to a continuous family labelled by the periods $(\nu_l, \mu_l)$ as in (\ref{periods}). 
\smallskip

Let us illustrate this for the simplest nontrivial case of $n=4$. The skeleton planar graph ${\cal G}$ is a tetrahedron with six edges, and so is the dual graph ${\cal G}^D$. The four constraints (\ref{nconstr}) at the vertices  of ${\cal G}$ imply that there are only two independent sets of $n_{ij}$.  We can take them to be, say, $n_{12}$ and $n_{13}$, and then all other $n_{ij}$ are determined in terms of these. The spectral curve determining the branched cover is of genus one, and thus there are four $a_j$ in eq.~(\ref{hypellip}), which correspond to the four vertices of ${\cal G}^D$. Generically there are therefore six period integrals taken along the six different cuts (or edges of ${\cal G}^D$), but only two of them are independent. If we choose these to be the cuts transverse to the edges $(12)$ and $(13)$ of ${\cal G}$, the corresponding periods in (\ref{periods}) are proportional to $n_{12}$ and $n_{13}$. Fixed values of these periods correspond to a particular graph with specified values of  $n_{12}$ and $n_{13}$, and each of them gives rise to a distinct covering map.

\section{The spectral curve and the Strebel differential}\label{sec:6}

In the previous section we have explained how the matrix model results are related to the Feynman diagrams of symmetric orbifold correlators that capture the different covering map contributions. In this section we want to show that, in the large twist limit, the sum over these discrete contributions becomes an integral over the moduli space of the covering space. The key observation that makes this possible is a remarkable relation between the spectral curve of the matrix model, and the Strebel differential on the moduli space of the covering space. As we will explain, our system therefore realises very explicitly the mechanism put forward some time ago in \cite{Gopakumar:2003ns, Gopakumar:2004qb, Gopakumar:2005fx} by means of which the string world-sheet path integral emerges from the dual CFT correlators.
\medskip

Recall from Section~\ref{sec:MM} that, to leading order in the large $N$ limit,  the spectral curve $y_0(z)$ has the form, see eqs.~(\ref{y0eq}) and (\ref{hypellip}) 
\be
y_0^2(z) = \frac{\alpha_n^2}{\prod_{i=1}^{n-1}(z-z_i)^2} \prod_{j=1}^{2n-4}(z-a_j) \ .
\ee
We also noted in Section~\ref{sec:4.3} that it agrees in the large $N$ limit with the Schwarzian of the covering map. Since the latter is a quadratic differential, see eq.~(\ref{SGtrans}), it follows that also
\begin{equation}\label{streb}
4\pi^2 \phi_S(z)dz^2 \equiv -y_0^2(z) dz^2 = -\frac{\alpha_n^2}{\prod_{i=1}^{n-1}(z-z_i)^2} \prod_{k=1}^{2n-4}(z-a_k) 
\end{equation}
defines a quadratic differential. It is clear from this explicit form that the quadratic differential $\phi_S(z)dz^2$  has double poles at $z_i$, as well as at $z_n=\infty$, and it follows from (\ref{spec-gen}) that the ``residues" at these double poles are $-\alpha_i^2$, which are therefore real and negative. We also see from eqs.~(\ref{hypellip}) and (\ref{y0eq}) that the only zeros are at $z=a_k$. Finally, the discussion of the previous subsection implies that all the periods around pairs of branch points $a_k$, see eq.~(\ref{periods}), are {\it real} and positive (with the appropriate orientation of the integral). 
 
These properties are precisely what characterises a \emph{Strebel differential} on the $n$-punc\-tu\-red sphere.\footnote{For a physicist friendly introduction to Strebel differentials see, for example, Section~3.2 of \cite{Mukhi:2003sz}, Section~3 of \cite{Gopakumar:2005fx}, Section~2 of \cite{Ashok:2006du}, or Appendix~A of \cite{Razamat:2008zr}.} 
Recall that at any point on the moduli space of $n$-punctured Riemann surfaces ${\cal M}_{g,n}$ there exists a Strebel differential, i.e.\ a unique meromorphic quadratic differential with only double poles (and specified negative residues) at the $n$ punctures, such that all the ``lengths" between zeroes are real,
\begin{equation}\label{streb-length}
l_{km}=\int_{a_k}^{a_m}\sqrt{\phi_S(z)} \in {\mathbb R}_+ \ .
\end{equation}
Each such Strebel differential $\phi_S(z)$  defines a critical graph on the Riemann surface, the so-called \emph{Strebel graph}, whose vertices are the zeros of the Strebel differential, and whose edges are the {\it critical horizontal trajectories}. Here {\it horizontal} means that the  curve $z(t)$ satisfies
\be\label{hortraj}
\phi_S\bigl(z(t)\bigr) \, \Bigl( \frac{dz}{dt} \Bigr)^2 > 0 \ ,
\ee
and a horizontal trajectory is {\it critical} if it is not closed, but rather connects two zeros of the Strebel differential, see Fig.~\ref{fig:1}.
These edges divide the Riemann surface into $n$ ring domains (faces with the topology of a disc), each of which contains exactly one double pole of $\phi_S$. This Strebel graph is therefore nothing other than ${\cal G}^D$ --- the dual to the skeleton graph ${\cal G}$ of the previous section.  
\medskip

The reason why this is significant is that Strebel differentials are, on the other hand, known to parameterise the (decorated) string moduli space. This is known as Strebel's Theorem \cite{Strebel}: for every Riemann surface $\Sigma_{g,n}$ with $n>0$ and $2g + n>2$, and any $n$ specified positive numbers $(p_1,\ldots,p_n)$, three exists a unique Strebel differential. This Strebel differential is holomorphic everywhere on $\Sigma_{g,n}$, except at the $n$ marked points where it has double poles with ``residue" equal to $-p_i^2$ at the $i$'th pole. 

As a consequence, each covering map is a contribution from a single point on moduli space since it is uniquely specified by the Strebel lengths in (\ref{periods}). Furthermore, the sum over all the branched covers defining the symmetric product correlator in (\ref{corrcov}) goes over, in the large twist limit, to an integral over the moduli space of the $n$-punctured sphere ${\cal M}_{0,n}$, where Strebel's Theorem guarantees that we cover the moduli space exactly once. 

Thus we have rewritten the symmetric orbifold correlators in the large twist limit as a world-sheet integral, with the world-sheet being the covering surface of the symmetric orbifold correlator. This therefore ties in very nicely with the proposal of \cite{Lunin:2000yv,Lunin:2001pw,Pakman:2009zz} that the covering surface should be identified with the world-sheet of the dual string theory; this was recently confirmed by an explicit world-sheet calculation \cite{Eberhardt:2019ywk,Dei:2020zui}. It also realises beautifully the general picture about how field theory diagrams combine into a world-sheet integral that was put forward by one of us some time ago \cite{Gopakumar:2003ns, Gopakumar:2004qb, Gopakumar:2005fx}.

\subsection[Finite \texorpdfstring{$N$}{N} generalisation]{\boldmath Finite \texorpdfstring{$N$}{N} generalisation}\label{sec:6.1}

We should note that the Strebel lengths are, in our large $N$ limit, proportional to the positive integers $n_{ij}$ which count the number of edges between vertices in the Feynman diagrams. This is somewhat reminiscent of the relation proposed by Razamat \cite{Razamat:2008zr} as an alternative to identifying the Schwinger parameters of the field theory \cite{Gopakumar:2005fx} with the Strebel lengths. Such a discrete relation was also seen to be very natural for the zero-dimensional Gaussian matrix model where one does not have any spacetime dependence in the correlators \cite{Razamat:2008zr, Razamat:2009mc}. In fact, the connection between Gaussian correlators and Belyi branched covering maps \cite{Koch:2010zza} (see also \cite{Itzykson:1990zb, DiFrancesco:1992cn}), and between the latter and integer length Strebel differentials \cite{mulpenk1}, made this relation compelling. It also suggested a candidate dual closed topological string theory \cite{Gopakumar:2011ev, Gopakumar:2012ny, Koch:2014hsa}.

This raises the natural question of what exactly the picture at finite $N$ should be in our case. For finite twist, there are only a finite number of covering maps, see e.g.\ eq.~(\ref{number}), and thus the symmetric orbifold correlator only gets contributions from isolated points in the moduli space, see also \cite{Eberhardt:2019ywk,Dei:2020zui}. It would be very interesting to understand what characterises the corresponding Strebel differentials. Note that this localisation fixes the 
 cross-ratios of the world-sheet coordinates $z_i$  in terms of those of the spacetime CFT, i.e.\ the $x_i$, and that the values of these cross ratios vary smoothly as we vary the $x_i$. In other words, the discrete points in the moduli space of the covering space (i.e.\ the world-sheet) are continuous functions of $x_i$, and cannot just be labelled by the discrete $n_{ij}$, unlike in the case of the Gaussian matrix model. The answer is presumably some discrete interpolation between the proper time prescription of \cite{Gopakumar:2005fx} and the integer length prescription of \cite{Razamat:2008zr}. We leave this important question for future investigation. 

It is also rather striking that to leading order in $N$, the Strebel differential is the same as  another natural quadratic differential one can associate to the covering map $\Gamma(z)$, namely the Schwarzian $S[\Gamma]$. In fact, we see from (\ref{y0schw}) that 
\be\label{strebschw}
4\pi^2 \phi_S(z)= -y_0^2(z) = \frac{2}{ N^2} S[\Gamma] - \frac{2}{N} W''(z) - \frac{4}{N} R_1(z) \ .
\ee  
Despite appearances, the first term on the RHS is actually of ${\cal O}(1)$ in the large $N$ limit, whereas the other two terms are down by factors of $\frac{1}{N}$. In fact, the  $\frac{1}{N}$ corrections are relatively mild: as described in Section~\ref{sec:4.3}, the second term on the RHS of (\ref{strebschw}) only corrects the residue of the double pole at $z=z_i$ from being proportional to $(w_i-1)^2$ to $(w_i^2-1)$, while the last term in (\ref{strebschw}) shifts the coefficient of the subleading simple pole at $z=z_i$. One may therefore think that, in some sense, the Strebel differential is essentially the Schwarzian of the covering map to all orders in $\frac{1}{N}$, although maybe not at finite $N$.

\section{Reconstructing the world-sheet}\label{sec:worldsheet}

In the previous section we have seen that one part of the program of \cite{Gopakumar:2004qb, Gopakumar:2005fx} is beautifully realised for the case of the symmetric orbifold correlators, at least in the limit of large twists: the sum of all Feynman diagram contributions to (free field) CFT correlators gives rise to an integral over the moduli space of Riemann surfaces of the dual string theory,
\be\label{schematic}
\sum_{\{n_{ij}\}} \ \ \longrightarrow \ \ \int \prod_{l=1}^{n-3}[d\nu_l d\mu_l]  =\int_{{\cal M}_{0,n}}|\omega^{(n-3)}(z_i)|^2 \ ,
\ee
where the flat measure in terms of the periods goes over to a top form in terms of the conventional $z_i$ parametrising the $n$-punctured sphere. Here we have used that the discrete sum over covering maps is indexed by the independent parameters $n_{ij}$, and that the sum over all such contributions goes over to an integral over the $(2n-6)$ independent periods or Strebel lengths in (\ref{periods}) in the large twist limit. The second equality uses the Jacobian of the transcendental relation between the Strebel lengths and the conventional moduli $z_i$, see \cite{David:2006qc} for a complete form of the relation for the $n=4$ case.
\medskip

The second part of the program of \cite{Gopakumar:2004qb, Gopakumar:2005fx} is to obtain the \emph{integrand} on moduli space from the dual CFT correlators, i.e.\ to reconstruct the actual world-sheet correlators from the spacetime perspective. As we have seen in (\ref{corrcov}) the spacetime CFT correlators are of the form 
\be\label{corrcov2}
\langle  \sigma_{w_1}(x_1) \cdots \sigma_{w_n}(x_n) \rangle = \sum_\Gamma W_{\Gamma} \prod_{i=1}^n |a_i^{\Gamma}|^{-2(h_i-h_i^0)}\, e^{-S_{\rm L}[{\Phi}_\Gamma]} \ ,
\ee
where the coefficients $a_i^{\Gamma}$ are defined via,
\begin{equation}
    \partial \Gamma(z)\sim a_{i}^{\Gamma} w_i \, (z-z_i)^{w_i-1} \ , \qquad  \hbox{as $z\to z_i$.}
\end{equation}
Here the sum over $\Gamma$ denotes the sum over all branched coverings, and this will become the integral over moduli space as in (\ref{schematic}). We are therefore interested in the large $N$ limit of the different summands in (\ref{corrcov2}).

Let us start with the first term, the one involving $a_i^\Gamma$. Using the form of $\partial \Gamma(z)$ in (\ref{partG}) and (\ref{covmap}), it follows that at leading order in $\frac{1}{N}$,
$a_i^{\Gamma}$ behaves as
\be
a_i^{\Gamma}=\frac{1}{w_i}\, M_{\Gamma}\,\prod_{j(\neq i)}^{n}(z_i-z_j)^{w_j-1}e^{-2N\,\int_{\cal C}d\lambda\, \rho(\lambda)\log (z_i-\lambda)} \ .
\ee
We also note that the factors of $(h_i-h_i^{0})\sim \mathcal{O}(1)$. Therefore the contribution from 
\be
\bigl|a_i^{\Gamma} \bigr|^{-2(h_i-h_i^0)} \sim e^{-N}\ .
\ee
This will turn out to be subdominant at large $N$ compared to the Liouville term which we analyse next. To evaluate the Liouville action, we note that the conformal factor ${\Phi}_{\Gamma}$ is given by (\ref{phisol})
\begin{equation*}
   {\Phi}_{\Gamma}(z,\Bar{z})=\ln(|\partial \Gamma|^2) \ ,
\end{equation*}
which implies, using (\ref{yGamma}), that 
\begin{equation}\label{phiy}
    \frac{1}{N}\partial  {\Phi}_{\Gamma} =  \frac{1}{N}\partial\log{\partial \Gamma} = y(z) \ , 
\end{equation}
where $y(z)$ is the spectral curve. To leading order in $N$, $y(z) \cong y_0(z) =i\sqrt{\phi_S(z)}$, and thus the classical Liouville action becomes, to leading order in $\frac{1}{N}$ 
\begin{equation}\label{Liouv}
{\rm S}_L[\Gamma] =  \, \frac{c}{48\pi} \int d^2z\, \Bigl( |\partial {\Phi}_{\Gamma}(z)|^2 
+ 2 R \Phi_\Gamma \Bigr)  
= \frac{cN^2}{48\pi} \int d^2z\, \Bigl( |\phi_S(z)| + \frac{2}{N^2} R \Phi_\Gamma \Bigr) \ .
\end{equation}
As explained in \cite{Lunin:2000yv} the $R\Phi_\Gamma$ term is only needed to regularise the contribution from infinity (in $z$-space), but does not otherwise contribute. If we ignore this regulator term the action is just given in terms of the Strebel differential, which defines 
an almost flat metric on the world-sheet with a line element given by 
\be\label{strebmet}
ds^2=|\phi_S(z)|dzd\bar{z} \rightarrow \sqrt{\det{g}} = |\phi_S(z)| \ .
\ee
Thus eq.~(\ref{Liouv}) is essentially the Nambu-Goto area action for the world-sheet in  ``Strebel gauge". Note that this Strebel gauge is characterised by the property that all its curvature is localised at the punctures or insertions of vertex operators, $z_i$, as well as at the zeroes $a_k$ of $\phi_S(z)$. The latter can be viewed as the interaction vertices of the string \cite{Gopakumar:2005fx}. This gauge had already appeared in the putative dual to the Gaussian model as was observed in \cite{Gopakumar:2012ny}. 

We should emphasise that the Strebel metric in (\ref{strebmet}) is  {\it distinct} from the induced metric on the covering space \cite{Lunin:2001pw}, which is given by the pullback  from the boundary ${\rm S}^2$ of the covering map 
\be\label{indmetr}
ds^2_{\text{pull}}= \big{|} \partial \Gamma(z) \big{|}^2 dz d\bar{z} \ .
\ee
The Strebel metric is instead to be thought of as an induced metric from the full dual ${\rm AdS}_3$ geometry. Indeed, the Liouville action (\ref{Liouv}) has been shown to arise as the classical on-shell action on ${\rm AdS}_3$ with the conformal factor ${\Phi}$ being identified with the radial direction \cite{Eberhardt:2019ywk}. From the ${\rm AdS}_3$ perspective, the relevant world-sheet is pinned to the insertion points at the boundary, but extends into the interior of ${\rm AdS}_3$, and  (\ref{Liouv}) should describe the `area' of this surface, i.e.\ the Strebel metric should be induced from this ${\rm AdS}$ embedding.\footnote{Since the ${\rm AdS}_3$ background also has a $B_{\rm NS}$-field, this will not just be the geometrical area. More generally, this picture also ties in with the general philosophy of \cite{Gopakumar:2004qb, Gopakumar:2005fx}, namely that it is a signature of the holographic nature of the dual world-sheet action.}  It would be very interesting to work this out in more detail. 

In this context it is very curious to note that there is also another way of viewing the on-shell Liouville action, which connects to old ideas of the rigid string \cite{Polyakov:1986cs, Kleinert:1986bk}. If we substitute (\ref{phiy}) in (\ref{Liouv}) and recall that the covering map $\Gamma(z)$ is nothing other than $X(z)$, which parametrises the boundary ${\rm S}^2$, then the Liouville action can be suggestively recast as (dropping as before the curvature term)
\be\label{rigid}
{\rm S}_L[X] =  \, \frac{c}{48\pi} \int d^2z\, \frac{1}{\partial X\bar{\partial}\bar{X}}\, \partial^2 X(z) \,\bar{\partial}^2\bar{X}(\bar{z}) \ .
\ee
This is an action for the purely two dimensional modes $X$, $\bar{X}$ of the large $N$ CFT. It is a four derivative action of the form very reminiscent of that which appears for rigid strings where one adds an extrinsic curvature term, compare with, say, eq.~(8) of \cite{Polyakov:1986cs}. Here, the curvature term may be some effective way of incorporating the extra (radial) dimension, and one would also have to incorporate the $B_{\rm NS}$-field that is ultimately responsible for the field $X(z)$ to be holomorphic. 
\medskip

Finally, there is yet another interesting way in which we can cast the Liouville action, again to leading order in $\frac{1}{N}$. Using (\ref{strebschw}) we can write, again dropping terms down by $\frac{1}{N}$ as well as the regulator term
\be\label{Liouv2}
      {\rm S}_L[\Gamma] =  \, \frac{c}{48\pi} \int d^2z\, |\partial {\Phi}_{\Gamma}(z)|^2 =   \frac{c}{24\pi }  \int d^2z\, \bigl| S[\Gamma](z)\bigr| \ . 
\ee      
This suggests a direct spacetime description somewhat analogous to what appears in the near ${\rm AdS}_2$ dual of the SYK models. Indeed, it was realised there that in terms of a bulk description such as the $JT$ gravity action, the Schwarzian action for the reparametrisation of the boundary ${\rm S}^1$ captures the low energy physics \cite{Maldacena:2016upp}. This can also be viewed as arising from a coadjoint orbit quantisation of the Virasoro group \cite{Mandal:2017thl}. On the other hand, the broken conformal symmetry of the SYK model \cite{Sachdev:1992fk, Kitaev-talk, Kitaev:2017awl}, also dictates a Schwarzian action for the low energy modes \cite{Maldacena:2016hyu}. In the present case, the presence of the correlators can be viewed as slightly breaking the 2d conformal symmetry and perhaps similar arguments can explain the universal nature of the Schwarzian action that governs the physics of the almost topological $k=1$ 
${\rm AdS}_3$ string theory.

\section{Discussion and outlook}\label{sec:conclusions}

Let us conclude by making a number of comments, and suggesting interesting directions for further research. To make this somewhat longish list of ideas more readable, we have organised the different points according to themes. 

\subsection*{The Gross-Mende like limit}

\begin{itemize}

\item
The large twist limit seems to be a fruitful regime to investigate correlators in the symmetric orbifold CFT, given how difficult it is to explicitly compute these even for small twist. It would be interesting, for instance, to study the four-point function specifically and connect with some of the techniques already employed in, for instance, \cite{Lunin:2000yv, Pakman:2009zz, Dei:2020}. Can we understand the limiting geometries of these covers better? Can we get a handle on systematic $\frac{1}{N}$ corrections?

\item
Another question is the extension of the considerations here to the case where the covering space/world-sheet is of higher genus. Is there a corresponding set of scattering equations and a role for a large $N$ matrix model reformulation? Gross-Mende saddle points in flat space, at higher genus, \cite{Gross:1987ar} were very simply related to those at genus zero. Is there also such a relation here? See also the discussion below on world-sheet correlators. 

\item
Relatedly, eq.~(\ref{scatt}) essentially gives rise to the flat space Gross-Mende equations if one also includes the poles $\lambda_a$ of the covering map as dynamical quantities. Is there a sense in which these are building up the world-sheet in our setup?

\item
We have looked at the correlators at fixed $x_i$. It would be interesting to look at Regge-like limits, perhaps in Mellin space. 

\item
Recently, a `large $p$' limit has been considered for half-BPS operators in 4d ${\cal N}=4$ Super Yang-Mills theory at strong coupling, with a similar aim of studying a Gross-Mende like limit  \cite{Aprile:2020luw}. The analogous scenario at weak coupling, informed by the approach of \cite{Gopakumar:2004qb, Gopakumar:2005fx}, is currently being  investigated \cite{wip}.

\end{itemize}

\subsection*{The Relation to Matrix Models}

\begin{itemize}

\item
We used matrix model technology to study the large $N$ limit of (\ref{scatt}). Is it possible to extend this to finite $N$ as well? Note that studying solutions of (\ref{scatt}) at finite $N$ is not the same as studying the finite $N$ matrix integral in (\ref{matint}). The loop equations (\ref{loopeqs}) and (\ref{yeq}), on the other hand, do hold at finite $N$, and we have taken some steps towards including the subleading effects. It would be good to do this more systematically and apply it to studying covering maps. In this context it is likely that the Schwarzian of the covering map, which naturally appears in the finite $N$ equations, see eq.~(\ref{schwloop}), will play a significant role.  

\item
Relatedly, the fact that the equations (\ref{scatt})  can be viewed as the saddle point equations for a matrix integral cries out for a deeper explanation. The fact that these are the same Penner-like potentials which appeared in the AGT context, in a triad with Seiberg-Witten theory and 2d Liouville CFT, might perhaps be a clue. Note that Strebel differentials have also made an appearance in Seiberg-Witten theory, see for example, \cite{Ashok:2006du, He:2015vua}.

\item
Even apart from the connection to Seiberg-Witten theory, one can ask the question whether these Penner-like matrix models have a meaning in the present context of the ${\rm AdS}_3/{\rm CFT}_2$ duality. 
Note that in the language of the matrix integral (\ref{matint}), the  potential (\ref{potential}) corresponds to insertions of powers of the operator ${\rm det}(z_i-M)$. 
These are suggestive of D-brane like operators, and perhaps one has an alternative open string description here along the lines of the open-closed-open string trialities of \cite{joburg-talk}, see also \cite{Bargheer:2019kxb, Jiang:2019xdz}. 

\end{itemize}

\subsection*{Feynman Diagrams and Strebel differentials}

\begin{itemize}

\item
We saw that a Strebel differential was naturally associated to the covering maps (and the corresponding Feynman diagrams) at large $N$. The corresponding Strebel lengths (\ref{periods}) took continuous real values signifying that one was covering all of moduli space. This leads to a measure on moduli space as in (\ref{schematic}). Can we get a handle on this top form in any natural geometric way? This should play an important role in understanding the world-sheet correlator.  

\item
This leads to the important question of how all of these considerations are modified for finite $N$ when we will only have a few discrete points on moduli space. What is the distribution of these points on moduli space? 
What characterises the Strebel differentials and the corresponding discrete Strebel lengths? In which way does it deform away from the large $N$ result of eq.~(\ref{periods})? Is the finite $N$ Strebel differential given by the Schwarzian of the covering map? The latter has all the right features but it is not clear if it has real periods. 

\item While finite $N$ is maybe too ambitious, it would also be interesting to consider the large $N$ limit, but to include all subleading $\frac{1}{N}$ corrections. Can we understand the structure of the Schwarzian of the covering map in this limit, and does this define a Strebel differential? 

\item
It is crucial for future generalisations to understand exactly the dictionary between the Strebel lengths and the position dependent field theory amplitudes, extending in some way the proposals of \cite{Gopakumar:2005fx, Razamat:2008zr}.
One possible way to obtain the relation between the Strebel lengths and the insertion points $x_i$, at finite $N$, is to see how this translates into a world-sheet OPE (along the lines of \cite{David:2006qc, Aharony:2007fs, David:2008iz}) which should then match that of the world-sheet correlators obtained in \cite{Eberhardt:2019ywk, Dei:2020zui}. More generally, it will be interesting to understand in more detail the connection this approach suggests between the world-sheet OPE and the spacetime OPE, see \cite{Aharony:2007rq, Ghosh:2017lti}. 

\end{itemize}

\subsection*{Reconstructing world-sheet correlators}

\begin{itemize}

\item 
In this paper we have found a very concrete realisation of the idea of \cite{Gopakumar:2004qb, Gopakumar:2005fx} for how to reconstruct the world-sheet theory describing strings on ${\rm AdS}$, from the dual conformal field theory perspective. However, for the case of ${\rm AdS}_3$ that is relevant for this paper, we actually knew already the world-sheet beforehand: the relevant string theory has minimal pure NS-NS flux, and it can be described in terms of an $\mathfrak{sl}(2,\mathds{R})$ (or $\mathfrak{psu}(1,1|2)$) WZW model \cite{Eberhardt:2018ouy,Eberhardt:2019ywk,Dei:2020zui}. It would therefore be interesting to confirm that the `reconstructed' world-sheet theory actually agrees with this WZW model. To a large extent this is already manifest --- in particular, the world-sheet correlators have exactly the expected form --- but there are some aspects that it would be interesting to check further. In particular, the analysis of Section~\ref{sec:worldsheet} allows one to reconstruct the weight $W_{\Gamma}$ with which the different world-sheet configurations contribute to the correlator, see eq.~(\ref{corrcov2}), and it would be interesting to rederive this from first principles from the WZW model perspective. (From that viewpoint, these weights should be fixed by crossing symmetry.) There is again an interesting interplay here between the spacetime CFT and the worldsheet CFT, and their respective bootstrap conditions \cite{caltech-talk}. 

\item
Continuing in this vein, it was already clear from \cite{Eberhardt:2019ywk} that the underlying Liouville action, see eq.~(\ref{Liouv}), agrees with the on-shell sigma model action for the exact semi-classical branched cover solutions of ${\rm AdS}_3$. However, in this paper we found different microscopic interpretations for it: we could either think of it as the Nambu-Goto action but in a special world-sheet gauge where the metric is that given by the Strebel differential.  Alternatively, we saw that we could view this action purely in terms of the boundary ${\rm S}^2$ --- the naive target space of the string theory --- where it takes a form similar to that of rigid strings \cite{Polyakov:1986cs, Kleinert:1986bk}. And finally, we could write it in terms of a Schwarzian action that is very reminiscent of what happens for the ${\rm AdS}_2$ duals of SYK models. It would be very interesting to understand these different viewpoints more directly. In each of these approaches, there is a signature of an extra dimension, and it would be very interesting to see how one can reverse engineer the entire {\it off-shell} ${\rm AdS}_3$ sigma model from these different viewpoints.

\item
In any of these different ways of phrasing the action to leading order in $\frac{1}{N}$, it would be very interesting to identify the relevant saddle point that dominates at large $N$. For example in the Schwarzian approach, this amounts to identifying the covering map which minimises the Schwarzian functional, see eq.~(\ref{Liouv2}), as we vary the $z_i$ over the $n$-punctured sphere, subject to the branching conditions at these insertions. Can we find this saddle point, and is there a nice geometric interpretation to it? Can we also understand the role of the finite $N$ corrections to this saddle point?  

\item
If we can find this Gross-Mende like saddle point for the genus zero covering space, we can ask about the corresponding saddles at higher genus. Are they related in some simple way through covers of the genus zero solution as in the flat space case \cite{Gross:1987ar}? Can we do a resummation of the pertubative expansion of the string theory in this limit? 

\item
The reader might have noticed that we have not used much information about the seed CFT whose symmetric product we are taking. We have also not used supersymmetry anywhere in our analysis. 
It is likely that there are further criteria that must be obeyed by the integrands on moduli space to be correlators of a {\it bona fide} string background. It will be important, going ahead, to understand what these are, and to see how they arise on the field theory side. 

\end{itemize}

\subsection*{Other Questions}

\begin{itemize}

\item
It might be timely to revisit the dual of the Gaussian matrix model along the lines of \cite{Razamat:2008zr, Gopakumar:2011ev}. In \cite{Gopakumar:2011ev, Gopakumar:2012ny}, based on the association of Feynman diagrams for the Gaussian correlators with Belyi maps \cite{Koch:2010zza}, a dual A-model closed topological string with target ${\mathbb P}^1$ was proposed. Recall that Belyi maps are also holomorphic covering maps of ${\mathbb P}^1$ with special branching, and thus the setting is very similar to the present one. In fact, in \cite{Gopakumar:2012ny, Koch:2014hsa} $n$-point correlators were compared on both sides,  with there being a precise matching in the limit of large operators, quite analogous to the large twist limit considered here. It would be instructive to see this matching at the level of world-sheet correlators like in our present example. 
 
\item
It would be very nice to use the present approach to get a more complete understanding of the string dual to 2d Yang-Mills theory \cite{Gross:1992tu, Gross:1993hu}. This, once again, is formulated in terms of branched covers of a target space by the world-sheet. The proposals of \cite{Cordes:1994sd, Horava:1993aq} predate holography, and it is likely that we can lift the topological rigid string theory of \cite{Horava:1993aq} to higher dimensions, as suggested by the considerations of Section~\ref{sec:worldsheet}. One obstacle to overcome is to generalise the considerations here to the partition function and non-local observables like Wilson loops. 

\item
Finally, to generalise these considerations to the all important case of four-dimen\-sio\-nal gauge theories, it would be fruitful to make a connection of this approach with the hexagon program and integrability, perhaps along the lines already put forward in \cite{Bargheer:2018jvq}.  

\end{itemize}

\acknowledgments

We thank Subhojoy Gupta, Shiraz Minwalla, Mahan Mj, and Spenta Wadia for useful conversations, and Ashoke Sen and Spenta Wadia for their comments on a draft of this paper. The work of MRG and BK was supported by the Swiss National Science Foundation through a personal grant and via the NCCR SwissMAP. The work of RG is supported in part by the J.C.~Bose Fellowship of the DST-SERB.
RG and PM acknowledge the support of the Department of Atomic Energy, Government
of India, under project no.~RTI4001, as well as the framework of support for the basic sciences by the people of India.

\appendix 

\section{Riemann-Hilbert solution to the Penner-like models}\label{app:RH}

In this appendix we present an alternative way of solving the matrix model to leading order in $\frac{1}{N}$. The idea is to solve the Riemann-Hilbert problem that is defined by the integral equation (\ref{intgr-eq})  
\begin{equation}\label{leading}
W^{\prime}(\lambda) = - \Bigl[ u(\lambda+i\epsilon)+u(\lambda-i\epsilon) \Bigr]  \qquad (\lambda \in {\cal C}) \ ,
\end{equation}
which we have rewritten in terms of the resolvent (\ref{wdef}). Note that the eigenvalue density itself is then given by the discontinuity of the resolvent to leading order in large $N$, see eq.~(\ref{evdens}). The general solution to this Riemann-Hilbert problem is of the form, see e.g.\ \cite{Marino:2004eq}
\begin{equation}\label{resolvent}
u_0(z)=\frac{1}{2}\oint_{\cal C} \frac{dv}{2\pi i} \frac{W'(v)}{z-v}\sqrt{\prod_{j=1}^{2 \ell}\frac{z-a_j}{v-a_j}} \ , 
\end{equation}
where $\ell$ is the number of cuts $[a_{2i-1},a_{2i}]$ with $i=1,\ldots,\ell$. 

For the case that is of interest to us, the 
potential $W(z)$ is given by eq.~(\ref{potential}). The integrand of the contour integral in (\ref{resolvent}) has poles at $v=z$, as well as at $v=z_i$, while there is no pole at infinity (since 
$W'(v)\sim \frac{1}{v}$). Carrying out the integral we therefore find 
\begin{equation}\label{res-sol}
u_0(z)=\frac{1}{2}\left[ W'(z)-\sum_{i=1}^{n-1}\frac{\alpha_i}{(z-z_i)}\sqrt{\prod_{k=1}^{2\ell}\frac{z-a_k}{z_i-a_k}}\, \right]  \ , 
\end{equation}
and the spectral curve (\ref{qtm-spec}) equals
\begin{equation}\label{spec-gen}
y_0(z)= \sum_{i=1}^{n-1}\frac{\alpha_i}{(z-z_i)}\sqrt{\prod_{k=1}^{2\ell}\frac{z-a_k}{z_i-a_k}} \ .
\end{equation}
The positions of the branch cuts, i.e.\ the $a_i$, are now determined by the requirement that the resolvent (\ref{res-sol}) goes as $\frac{1}{z}$ as $z\to\infty$. This is not obvious since the second term in the bracket goes as $\sim z^{\ell -1} + \cdots$. Requiring that these higher powers of $z$ vanish (and that the coefficient of the $z^{-1}$ is equal to $1$), leads to the $\ell+1$ equations 
\begin{equation}\label{res-cond}
\sum_{i=1}^{n-1}\tilde{\alpha}_i z_i^m= \alpha_0\, \delta_{m, \ell} \ ,  \qquad m=0,\ldots, \ell \ ,
\end{equation}
where we have defined 
\begin{equation}
\tilde{\alpha}_i =\frac{\alpha_i}{\sqrt{\prod_{k=1}^{2\ell}(z_i-a_k)}} \ , \qquad \hbox{and} \qquad 
\alpha_0=\sum_{i=1}^{n-1}\alpha_i -2 = -\alpha_n - \frac{2}{N} \ , 
\end{equation}
and we have used eq.~(\ref{sumalph}) for the final equation. We would expect a maximum of $\ell = n-2$ cuts because there are $(n-2)$ critical points of $W$, $W'(z^\ast_j)=0$, $j=1,\ldots,n-2$, and usually the branch cuts emerge from the different critical points. 
For $\ell=n-2$, the system of $\ell+1$ homogeneous equations in (\ref{res-cond}) is invertible 
$\tilde{\alpha}_i$,
\be\label{reseqs}
\tilde{\alpha}_i = \frac{\alpha_i}{\sqrt{\prod_{k=1}^{2n-4}(z_i-a_k)}} = \frac{a}{\prod_{j\neq i} (z_i - z_j)} \ , \qquad i = 1,\ldots, (n-1) \ , 
\ee
where $a$ is a constant. Plugging this solution into eq.~(\ref{spec-gen}) the spectral curve then simplifies to 
\be\label{spec-fin}
y_0(z) = \frac{\alpha_n}{\prod_{i=1}^{n-1} (z-z_i)} \, \sqrt{\prod_{k=1}^{2n-4} (z-a_k)} \ . 
\ee
Here we have used the fact that the residue at $z_n=\infty$ is $\alpha_n$. We see that we have arrived at the same leading order spectral curve as we did through the  loop equations --- see eqs.~(\ref{y0eq}) and (\ref{hypellip}). 

Note that in this way of solving the matrix model, we use the $(n-1)$ resolvent equations in (\ref{res-cond}), as well as the $2(n-3)$ independent period equations of the spectral curve --- over both the A- and B-cycles --- in (\ref{periods}). These are then $(3n-7)$ conditions for the $(2n-4)$ $a_k$'s, as well as the $(n-3)$ cross ratios of the $z_i$. (Recall we have used the M\"obius invariance to fix three of the $z_i$). Thus the counting also works out as expected.

\section{Deriving the loop equations}\label{loop-eq}

In this appendix we derive the loop equations (\ref{loopeqs}). 
We begin with the scattering equations rewritten as (\ref{scatt2})
\begin{equation}
\frac{1}{2} W'(\lambda_a) = \frac{1}{N} \sum_{b\neq a} \frac{1}{\lambda_a - \lambda_b} \ . 
\end{equation}
Following \cite{Marino:2004eq}, we now multiply both sides by $\frac{1}{(\lambda_a-z)}$ and sum over $a$ to obtain
\be\label{Sdef}
\frac{1}{2} \sum_{a=1}^{N} \frac{W'(\lambda_a)}{(\lambda_a-z)} = \frac{1}{N} \sum_{a\neq b} \frac{1}{(\lambda_a - \lambda_b)(\lambda_a-z)} \equiv {\cal S}\ . 
\ee
We can rewrite ${\cal S}$ in terms of partial fractions as 
\be
{\cal S} = \frac{1}{N} \sum_{a\neq b} \frac{1}{(\lambda_a - \lambda_b)(\lambda_a-z)} = \frac{1}{N} \sum_{a\neq b} \frac{1}{(\lambda_b - z)} \Bigl[ \frac{1}{(\lambda_a-\lambda_b)} - \frac{1}{(\lambda_a-z)} \Bigr] \ . 
\ee
The first term on the right-hand-side equals $- {\cal S}$ --- it is obtained from ${\cal S}$ in (\ref{Sdef}) upon exchanging the dummy variables $a\leftrightarrow b$ --- and thus we deduce that 
\begin{eqnarray}
{\cal S} & = &  - \frac{1}{2N} \sum_{a\neq b} \frac{1}{(\lambda_a - z)(\lambda_b - z)} = - \frac{1}{2N} \Bigl( \sum_{a=1}^{N} \frac{1}{(z-\lambda_a)} \Bigr)^2 + \frac{1}{2N} \sum_{a=1}^{N} \frac{1}{(z-\lambda_a)^2} \nonumber \\
& = & - \frac{N}{2} u^2(z)  - \frac{1}{2} u'(z) \ , 
\end{eqnarray}
where $u(z)$ is the resolvent of eq.~(\ref{wdef}), which for finite $N$ becomes
\be
u(z) = \frac{1}{N} \sum_{a=1}^{N} \frac{1}{(z-\lambda_a)} \ . 
\ee 
Plugging this back into eq.~(\ref{Sdef}) we therefore deduce that 
\be\label{step1}
\frac{1}{2N} \sum_{a=1}^{N} \frac{W'(\lambda_a)}{\lambda_a-z} = - \frac{1}{2}u^2(z) - \frac{1}{2N} u'(z) \ . 
\ee
If we introduce the function $R(z)$ via 
\be
R(z) = \frac{1}{N} \sum_{a=1}^{N} \frac{ W'(\lambda_a) - W'(z)}{(\lambda_a - z)} \ , 
\ee
then (\ref{step1}) becomes 
\be\label{step2}
  u^2(z) -  W'(z) u(z) +  \frac{1}{N} u'(z) +  R(z) = 0 \ ,
\ee
see eq.~(\ref{loopeqs}). We note that for our Penner like potential 
\begin{equation}\label{Req1}
R(z) = \frac{1}{N}\sum_{a=1}^N\frac{W^{\prime}(z)-W^{\prime}(\lambda_a)}{z-\lambda_a} =
-\frac{1}{N}\sum_{i=1}^{n-1}\frac{\alpha_i}{z-z_i} \sum_a\frac{1}{\lambda_a-z_i}= \sum_{i=1}^{n-1}\frac{\alpha_iu(z_i)}{z-z_i} \ ,
\end{equation}
where we have used the definition (\ref{wdef}) for the resolvent. We also note that 
\begin{equation}\label{wconstr}
\sum_{i=1}^{n-1} \alpha_iu(z_i) = -\frac{1}{N}\sum_{i=1}^{n-1}\sum_a\frac{\alpha_i}{\lambda_a-z_i}
= - \frac{2}{N^2}\sum_{a} \sum_{b\neq a} \frac{1}{\lambda_a-\lambda_b} =0 \ , 
\end{equation}
where we have used the saddle point equation (\ref{scatt2}) in the second last equality, and antisymmetry under $(a,b)$ exchange in the last. 
These relations again hold at finite $N$.

\end{document}